\documentclass[useAMS, usenatbib]{mn2e}

\newcommand{\eqref}[1]{(\ref{#1})}

\usepackage{graphicx}

\title[Theoretical Uncertainties due to AGN Subgrid Models]
{Theoretical Uncertainties due to AGN Subgrid Models in Predictions of Galaxy Cluster Observable Properties}

\author[H.-Y.\ K.\ Yang et al.]
{H.-Y.\ Karen Yang,$^1$\thanks{Email: hsyang@umich.edu},
P.\ M.\ Sutter$^2$ 
and Paul M.\ Ricker$^{3,4}$ \vspace{8pt} \\
$^1$Department of Astronomy, University of Michigan, Ann Arbor, MI\vspace{5pt}\\
$^2$Department of Physics, University of Illinois, Urbana, IL\vspace{5pt}\\
$^3$Department of Astronomy, University of Illinois, Urbana, IL\vspace{5pt}\\
$^4$National Center for Supercomputing Applications, Urbana, IL}

\date{Submitted 2011 December}

\pagerange{\pageref{firstpage}--\pageref{lastpage}} \pubyear{2011}

\def\LaTeX{L\kern-.36em\raise.3ex\hbox{a}\kern-.15em
    T\kern-.1667em\lower.7ex\hbox{E}\kern-.125emX}

\begin{document}

\label{firstpage}

\maketitle

\begin{abstract}

Cosmological constraints derived from galaxy clusters rely on accurate predictions of cluster observable properties, in which feedback from active galactic nuclei (AGN) is a critical component. In order to model the physical effects due to supermassive black holes (SMBH) on cosmological scales, subgrid modeling is required, and a variety of implementations have been developed in the literature. However, theoretical uncertainties due to model and parameter variations are not yet well understood, limiting the predictive power of simulations including AGN feedback. By performing a detailed parameter sensitivity study in a single cluster using several commonly-adopted AGN accretion and feedback models with FLASH, we quantify the model uncertainties in predictions of cluster integrated properties. We find that quantities that are more sensitive to gas density have larger uncertainties ($\sim 20\%$ for $M_{\rm gas}$ and a factor of $\sim 2$ for $L_X$ at $R_{500}$), whereas $T_X$, $Y_{SZ}$, and $Y_X$ are more robust ($\sim 10-20\%$ at $R_{500}$). To make predictions beyond this level of accuracy would require more constraints on the most relevant parameters: the accretion model, mechanical heating efficiency, and size of feedback region. By studying the impact of AGN feedback on the scaling relations, we find that an anti-correlation exists between $M_{\rm gas}$ and $T_X$, which is another reason why $Y_{SZ}$ and $Y_X$ are excellent mass proxies. This anti-correlation also implies that AGN feedback is likely to be an important source of intrinsic scatter in the $M_{\rm gas}$--$T_X$ and $L_X$--$T_X$ relations.

\end{abstract}

\begin{keywords}
galaxies: clusters: general --- hydrodynamics --- methods: numerical --- galaxies: active
\end{keywords}

%===================================================

\section{Introduction}

Clusters of galaxies are useful probes of cosmological parameters, provided that their masses can be determined accurately from multi-wavelength observations calibrated using theoretical models. However, it is still a challenge for current theoretical models to reproduce all the observed properties of the baryonic content of clusters. Despite the fact that current cosmological simulations with radiative cooling and supernova feedback are able to reproduce profiles of the intracluster medium (ICM) outside the cores, the simulated cluster cores generally suffer from the over-cooling problem, e.g., the fraction of cool-core (CC) clusters and stellar fraction in these simulations are too high compared to observed values (see review by \citet{Borgani09} and references therein). Therefore, some additional forms of heating are required to suppress cooling. Feedback from active galactic nuclei (AGN) is one of the most promising candidates, as observations of X-ray cavities blown by jets from the central AGN imply a mechanical power that is comparable to the X-ray luminosity, suggesting a feedback loop might be at work \citep{Dunn08}. 

Since a wide range of length scales is involved in this feedback loop, from the accretion disk of the supermassive black hole (SMBH) on parsec scales to clusters on Mpc scales, direct simulation with all relevant physics is beyond current computational power. Thus cosmological simulations with AGN feedback to date have had to model its sub-resolution effects by linking the resolvable scale (usually $\sim$kpc) to the SMBH accretion disk scale using some simplified assumptions \citep{Sijacki, Booth, Dubois, Gaspari}. These studies have had success in explaining the cosmic evolution of SMBH, star formation history, and local scaling relations between the black hole mass and host properties, though current AGN models are still quite phenomenological. One important question to ask is whether these simulations can simultaneously reproduce observed ICM properties as well. Analyses in this direction have started recently \citep{Puchwein08, Gaspari11} and need to be further addressed. 

If cosmological simulations with AGN could accurately predict the core properties of clusters, they could provide crucial information for cluster cosmology, such as the CC fraction as a function of mass and redshift, or the parameterizations of scaling relations. The CC fraction is important for understanding the X-ray selection bias toward CC clusters because of their peaked central surface brightness. For calibrating the scaling relations of multi-wavelength cluster surveys, one may not be comparing apples to apples if such a selection bias is not adequately accounted for. Moreover, the core properties may have an impact on the evolution in the slopes, normalizations, and scatter of the scaling relations. In order to obtain meaningful cosmological constraints using the self-calibration method \citep{Levine02, Majumdar04, Lima04}, the parametrizations of the scaling relations have to be informed by numerical simulations.

However, the uncertainties in the existing AGN subgrid models are not yet well understood. Since it is still unknown how to link the accretion and feedback across different scales, there is a great amount of freedom to implement and parametrize the AGN subgrid models. The model parameters sometimes do not have a clear connection to observable quantities, and hence constraints from observations cannot be easily applied. Moreover, because these cosmological simulations require significant computational resources to run, it is difficult to perform detailed parameter studies to assess the robustness of the results. But in order to achieve predictions with high precision and controlled systematics, it is necessary to properly parametrize our ignorance in the AGN models and quantify the theoretical uncertainties. 

The aim of this study is to quantify the current theoretical uncertainties due to model variations in predicting the global ICM properties and thus provide a general guideline for cosmological simulations including AGN feedback. To this end, we implement a subgrid AGN unit in the FLASH simulation code that incorporates several existing AGN accretion and feedback models. To study the effect of AGN feedback on cluster observables, we put these models in an idealized cluster and explore a wide range of parameters systematically. Connections between the model parameters and observable quantities are provided whenever possible. We identify the numerical details and parameters that the results are most sensitive to. For all the models that successfully self-regulate black hole growth and reproduce observed cluster profiles, we then study the influence of AGN feedback on integrated cluster properties.  

The structure of this paper is as follows. The analytical and numerical approaches are described in \S~\ref{sec:method}. The result of the sensitivity test is presented in \S~\ref{sec:sensitivity}. In \S~\ref{sec:observables}, we first quantify the model uncertainties of integrated cluster properties, and then study the effects of AGN feedback on the scaling relations. Finally, we conclude and discuss possible ways to improve the subgrid models in \S~\ref{sec:conclusion}.

%===================================================

\section{Methodology}
\label{sec:method}

\subsection{Simulation Setup}

We performed three-dimensional hydrodynamic simulations with radiative cooling and AGN feedback within an isolated cluster sitting in a 2048\ kpc box using the adaptive mesh refinement (AMR) code FLASH\ 3 \citep{Flash, Dubey08}. The cluster is set up in the same way as in \citet{Cattaneo} and has properties similar to M87. The cluster gas is initialized assuming a polytropic equation of state (EOS) \citep{KS01} and is in hydrostatic equilibrium in a fixed NFW \citep{NFW} gravitational potential. Self-gravity of the gas is not included. The cluster has a virial mass of $1.5\times 10^{14}\ \mbox{M}_\odot$, concentration of 5.53, and gas fraction of 0.1. This gas fraction is chosen to match the observed value for clusters of similar masses \citep[e.g.][]{Gonzalez07}. The contribution to the baryon fraction due to stellar mass is not included because it is small compared to the gas contribution. A $3\times 10^9\ \mbox{M}_\odot$ black hole (BH) is placed in the center; it is only used for computing the accretion and feedback quantities and does not contribute to the gravitational potential. The base grid resolution is 32\ kpc throughout the simulation volume, and the region surrounding the central black hole is refined progressively to the maximum resolution (e.g.\ 1.0 kpc for the fiducial run). To accommodate the extent of feedback with high resolution, we define the maximally-refined region as a box of width 120\ kpc for the bubble feedback model and 60\ kpc for the jet model (see \S~\ref{sec:feedback_model} for details of these models). The diode boundary condition is used; this is similar to the outflow boundary condition but does not allow matter to flow into the domain. 

Radiative cooling is computed using \citet{SutherlandDopita} assuming 1/3 solar metallicity. Star formation and feedback from supernovae are neglected because they themselves have different implementations and require detailed comparisons of their own \citep[e.g.][]{McCarthy11}. In this study we intend to investigate the modeling of AGN alone and avoid confusion due to possible interference with other subgrid physics. We do not expect our main conclusions to change because in most runs the gas densities never reach the conventional star formation threshold of $n_H=0.1\ \mbox{cm}^{-3}$. Also, in our analysis of cluster integrated properties, we choose observables that are insensitive to the dense, cold gas surrounding the SMBH (see \S~\ref{sec:observables}). Note also that in this paper we only focus on the hydrodynamic models employed in previous cosmological simulations, and hence the effects of magnetic fields are neglected. We refer readers to \citet{Sutter12} for an investigation of magnetized AGN feedback models. 
The Hubble constant $h=0.65$ is used. When overdensity quantities are quoted, they are computed using the overdensity radius $R_\Delta$ where the enclosed average density is $\Delta$ times the critical density of the universe. 

%-------------------------------------------------------------------------------------------

\subsection{AGN Subgrid Models}

Subgrid models in cosmological simulations have been developed with varying levels of sophistication. Accretion rate calculations can vary from the simple Bondi \citep{Bondi} rate or its modifications \citep{Sijacki, Booth}, to accretion of cold gas only \citep{Pizzolato05, Gaspari11}. To model quasars at high redshifts, it is commonly assumed that a fraction of their radiative energy is transformed into thermal energy in the surrounding gas \citep[e.g.][]{DiMatteo05, Bhattacharya07}. The mechanical input from AGN can be modeled using large-scale jets when the resolution permits \citep[e.g.][]{Cattaneo, Sternberg07, Dubois, Morsony10, Gaspari}. Since jets eventually inflate bubbles, it is easier computationally to place already-formed bubbles \citep[e.g.][]{Sijacki, Booth}.   

Since it is infeasible for us to explore all the models presented above, we select some representative accretion and feedback models. For estimating the BH accretion rate, we consider the $\alpha$ model proposed by \citet{Sijacki}. The feedback from the AGN is modeled based on the bubble feedback of \citet{Sijacki} or the jet feedback of \citet{Cattaneo}. Since our main goal is to quantify the model uncertainties, these models are chosen to cover very different methods of implementation and parameterization. We summarize the important aspects of these models in the following. 

\subsubsection{SMBH Accretion}

To include the accretion onto the SMBH self-consistently in cosmological simulations, the simplest approach is to estimate the accretion rate using the Bondi-Hoyle-Lyttleton \citep{Bondi} accretion rate:
\begin{equation}
\dot{M}_{\mathrm{BH}} \propto \dot{M}_{\mathrm{Bondi}} = 4\pi G^2 M^2_{\mathrm{BH}} \rho / c_{\mathrm{s}}^3,
\end{equation}
where $M_{\mathrm{BH}}$ is the BH mass, and $\rho$ and $c_{\mathrm{s}}$ are the gas density and sound speed, respectively. Cosmological simulations usually do not have sufficient resolution to resolve the Bondi radius, $r_{\rm Bondi} \equiv GM/c_s^2$, as well as the multiphase gas when the density is high enough to trigger star formation. Therefore the density (temperature) at the Bondi radius would likely be higher (lower) than values on the grid. The actual Bondi accretion rate would thus be underestimated, which is reflected by the proportionality in the above equation. 

The $\alpha$ model assumes a constant coefficient, i.e., $\dot{M}_{\mathrm{BH}} = \alpha \dot{M}_{\mathrm{Bondi}}$. \footnote{It is possible to consider other forms of proportionality, such as the $\beta$ model proposed by \citet{Booth}, which is consistent with the Bondi prediction when simulations have sufficient resolution or when gas densities are lower than the star formation threshold $n_H=0.1\ \mbox{cm}^{-3}$; otherwise the proportionality is density dependent to account for accretion of multiphase gas. Since in our simulations the gas densities seldom reach the star formation threshold, the $\beta$ model is equivalent to $\alpha=1$ in our current setup.} Based on the resolution argument, $\alpha = 1$ is justified for our simulated cluster because of its flat gas profile. However, the value of $\alpha$ is often taken to be $\sim 100$ in previous works to drive substantial black hole growth \citep{DiMatteo05, Sijacki, Bhattacharya07}, indicating the large uncertainty in the estimation of accretion rates. In our parameter survey, we will vary $\alpha$ from 1 to 100 in order to cover situations where the actual accretion rate is underestimated as well as where it is overestimated.

After computing the accretion rate, an upper limit is imposed corresponding to the Eddington rate, $\dot{M}_{\mathrm{Edd}} = (4\pi G M_{\mathrm{BH}} m_{\mathrm{p}}) / (\epsilon_{\mathrm{f}} \sigma_{\mathrm{T}} c)$, where $m_{\mathrm{p}}$ is the mass of the proton, $\sigma_{\mathrm{T}}$ is the Thompson cross-section and $\epsilon_{\mathrm{f}} $ is the radiative efficiency.  

The region for computing the accretion rate and the region from which to deplete the accreted gas are typically set to a few zones around the central BH, but their sizes are essentially arbitrary. We denote these two radii as $R_\mathrm{acc}$ and $R_\mathrm{dep}$, and we probe several values in \S~\ref{sec:numerical}. 
Note that during strong accretion events, gas in a grid cell can possibly be completely removed and cause an unphysical surge of gas around the black hole. In such cases, we increase the depletion radius by a small amount so that no more than $10\%$ of the gas on a grid cell is removed in one timestep. We expect this condition to have negligible effects because it does not occur frequently, and the results are insensitive to the depletion radius (see \S~\ref{sec:numerical}).

\subsubsection{AGN Feedback}
\label{sec:feedback_model}

The feedback from the AGN to the surrounding gas is then computed according to the accretion rate. There has been growing observational evidence for an anti-correlation between radio loudness and SMBH accretion rate \citep{Ho02, Sikora07}. That is, radio jets are associated with systems having lower accretion rates, while objects with higher accretion rates, like quasars at higher redshifts, are radiatively efficient, analogous to states of X-ray binaries \citep{Fender99, Gallo03}. For this reason, we follow the prescription in \citet{Sijacki} for switching to the quasar mode when the accretion reaches 1\% of the Eddington rate. In the quasar mode, the radiative energy is thermally coupled to the surrounding gas: the energy deposition rate is $\dot{E} = \epsilon_{\mathrm{r}} \epsilon_{\mathrm{f}} \dot{M}_{BH} c^2$, where $\epsilon_{\mathrm{r}}$ is the quasar heating efficiency. The region into which to dump the quasar thermal energy is chosen to be four zones in radius, though it is arbitrary. Note that in some other subgrid models \citep[e.g.][]{Booth} there is no division between the quasar and mechanical feedback since they are both assumed to be purely thermal and spherically distributed. However, we will show that when feedback energy is injected in the thermal form, the size of the region has a significant impact on the results.    

At low accretion rates, one can choose either bubble or jet feedback. For the bubble feedback \citep{Sijacki}, the feedback energy is distributed in terms of thermal energy within a spherical region around the SMBH. Bubbles are only formed when the BH mass increases by a fraction $\delta_{\mathrm{BH}}$ since the last bubble formation. When a bubble is formed, purely thermal energy is injected:
\begin{equation}
E = \epsilon_{\mathrm{m}} \epsilon_{\mathrm{f}} c^2 \delta M_{\mathrm{BH}},
\label{eq:ebub}
\end{equation}
where $\epsilon_{\mathrm{m}}$ is the efficiency of mechanical heating, and $\delta M_{\mathrm{BH}} \equiv \delta_{\mathrm{BH}} M_{\mathrm{BH}}$ is the increase in BH mass since the last bubble was formed. The injected energy is distributed in a mass-weighted sense within a sphere of radius
\begin{equation}
R = R_0 \left( \frac{E/E_0}{\rho/\rho_0} \right)^{1/5}, 
\label{eq:rbub}
\end{equation}
where the scaling parameter values $R_0$, $E_0$, and $\rho_0$ are motivated by observed bubble sizes. The bubble centers are randomly displaced within a sphere of radius $R_{\rm dis}$ centered on the black hole. In the fiducial run $R_{\rm dis}=R$. We also experiment with cases where bubbles are fixed at the central black hole, i.e., $R_{\rm dis}=0$.

In contrast to bubble feedback, which only injects thermal energy into the surrounding ICM, the jet feedback simulations inject some or all energy in kinetic form. The jet models are not intended to simulate the relativistic jet directly, but rather the non-relativistic outflow from the accretion disk \citep{Proga07} or decelerated large-scale jet after it has entrained some intergalactic medium during its propagation \citep{Feretti99, Laing02}. The ratio of injected thermal to kinetic energy depends on the parametrization and is different from study to study. While \citet{Gaspari} and \citet{Dubois} adopted purely kinetic jets, in \citet{Cattaneo} the injected energy is mostly thermal, depending on the amount of mass loading. The model of \citet{Cattaneo} is motivated by the observation that more massive, slow jets should couple more thermal energy with the surroundings as they propagate. However, a purely kinetic jet in their model would produce relativistic velocities, which cannot be treated adequately in non-relativistic hydrodynamic simulations. Therefore, we modified their model and used a more general parametrization to allow tuning of the thermal to kinetic ratio, while the jet velocities are independently determined by the amount of mass loading.   

Our generalized jet model can be summarized as follows. The injection rates of mass, momentum, and energy onto the grid are treated as source terms in the hydrodynamic equations and are calculated by
\begin{eqnarray}
\dot{M} &=& \eta \dot{M}_{\mathrm{BH}} |\Psi|, \nonumber \\
\dot{P} &=& \sqrt{2\eta \epsilon_{\mathrm{f}} (1-\epsilon_{\mathrm{m}}) } \dot{M}_{\mathrm{BH}} c \Psi, \label{eq:jets}\\
\dot{E} &=& \epsilon_{\mathrm{f}} \dot{M}_{\mathrm{BH}} c^2 |\Psi|, \nonumber
\end{eqnarray}
where $\eta$ is the mass loading factor and $E$ is the sum of injected thermal and kinetic energy, i.e., $E=E_{\rm th} + E_{\rm k}$. For non-relativistic jets, $E_{\rm k} = P^2/2M$; thus in this model $\epsilon_{\mathrm{m}}$ of the total energy goes into the thermal energy, and the remainder is kinetic. The jet velocity is 
$c\sqrt{2\epsilon_{\mathrm{f}} (1-\epsilon_{\mathrm{m}})/\eta}$, which is $\sim 10^4$ km/s for $\eta=100$, $\epsilon_{\mathrm{f}} = 0.1$ and $\epsilon_{\mathrm{m}} = 0$.
The function $\Psi$ determines the spatial extent of the jet:
\begin{equation}
\Psi ({\bf x}) = \frac{1}{2\pi r^2_{\mathrm{ej}}} \exp \left( - \frac{x^2+y^2}{2r^2_{\mathrm{ej}}} \right) \frac{z}{h^2_{\mathrm{ej}}}.
\label{eq:psi}
\end{equation}
The jet is aligned with the $z$-axis, and the feedback is applied to regions with $|z| \leq h_{\mathrm{eq}}$ and $\sqrt{x^2 + y^2} \leq r_{\mathrm{ej}}$. We also normalize the window function $\Psi$ so that the total injected energy in the cylinder sums up to $E$. Note that there is no threshold for injecting the jets, so the jet feedback is continuous rather than episodic.  

Note that our jet model includes several modifications to that in \citet{Cattaneo}. In addition to changes in the parametrization as described above, we normalized the function $\Psi$ as in their subsequent papers based on the same model \citep{Dubois}. They fixed the mass of the black hole for computing the accretion rate, while our black hole can grow from gas accretion. This has negligible effects since the accretion rate is small in their setup. Also, we allow gas depletion and use a smaller radius for computing the accretion rate. When accounting for the above differences, we are able to reproduce their results.  

%-------------------------------------------------------------------------------------------

\subsection{Model and Parameter Variations}
\label{sec:models}

Since the details of each subgrid model are so different that it is infeasible to explore every implementation and parameter, in this study we focus on those aspects of these models which are least constrained. To this end, the bubble and jet feedback models are chosen because they are different in many aspects, including the form of injected energy (thermal vs.\ kinetic), shape of injection region (spherical vs.\ jet-like), and periodicity of feedback (episodic vs.\ continuous). Comparing these two distinct models allows us to understand the extent of current theoretical uncertainties due to these AGN models. 

\begin{table}%[tp]
\caption{Survey of Numerical Parameters in the Bubble Model.}
\begin{center}
\begin{tabular*}{0.25\textwidth}{@{\extracolsep{\fill}} ccc}
\hline
\hline
Name & $\Delta x$ (kpc) & $\alpha$ \\ % \footnotemark[1]\\ 
\hline
&&\\[-3pt]
\multicolumn{3}{c}{Varying Resolution} \\
\hline
B1A & {\bf 0.5} & 1 \\
B1B & {\bf 1.0} & 1 \\
B1C & {\bf 2.0} & 1 \\
B1D & {\bf 4.0} & 1 \\
B1E & {\bf 8.0} & 1 \\
&&\\[-3pt]
\multicolumn{3}{c}{Scaling Alpha with Resolution}\\
\hline
B2A & {\bf 2.0} & {\bf 2} \\
%B2B & 4.0 & 300 \\
\hline
\hline
\end{tabular*}
\end{center}
\label{tbl:num_bub_par}
\end{table}

\begin{table}%[tp]
\caption{Survey of Numerical Parameters in the Jet Model.}
\begin{center}
\begin{tabular*}{0.45\textwidth}{@{\extracolsep{\fill}} cccccc}
\hline
\hline
Name & $\Delta x$ (kpc) & $h_\mathrm{ej}$ & $r_\mathrm{ej}$ (kpc) & $R_\mathrm{acc}$ & $R_\mathrm{dep}$ (zones) \\ % \footnotemark[1]\\ 
\hline
&&&&&\\[-3pt]
\multicolumn{6}{c}{Varying Resolution}\\
\hline
J1A & {\bf 0.5} & 2.0 & 2.5 & 2 & 2 \\
J1B & {\bf 1.0} & 2.0 & 2.5 & 2 & 2 \\
J1C & {\bf 2.0} & 2.0 & 2.5 & 2 & 2 \\
&&&&&\\[-3pt]
\multicolumn{6}{c}{Scaling Jet Size with Resolution} \\
\hline
J2A & {\bf 1.0} & {\bf 4.0} & {\bf 5.0} & 2 & 2 \\
J2B & {\bf 2.0} & {\bf 8.0} & {\bf 10.0} & 2 & 2 \\
J2C & {\bf 4.0} & {\bf 16.0} & {\bf 20.0} & 2 & 2 \\
J2D & {\bf 1.0} & {\bf 8.0} & {\bf 10.0} & 2 & 2 \\
&&&&&\\[-3pt]
\multicolumn{6}{c}{Varying Radii for Accretion and Depletion}\\
\hline
J3A & 1.0 & 2.0 & 2.5 & {\bf 1} & {\bf 1} \\
J3B & 1.0 & 2.0 & 2.5 & {\bf 4} & {\bf 4} \\
J3C & 1.0 & 2.0 & 2.5 & {\bf 2} & {\bf 0} \\
\hline
\hline
\end{tabular*}
\end{center}
\label{tbl:num_jet_par}
\end{table}

\begin{table*}%[tdp]
\caption{Survey of Physical Parameters.}
\begin{center}
\begin{tabular*}{0.9\textwidth}{@{\extracolsep{\fill}} ccccccc}
\hline
\hline
Name & $\alpha$ & Feedback & $\epsilon_\mathrm{f}$ & $\epsilon_\mathrm{m}$ & $\delta_{\mathrm{BH}} (\%)$ & Region \\ 
\hline
&&\\[-3pt]
\multicolumn{7}{c}{Varying Accretion}\\
\hline
P1A & {\bf 1} & bubble & 0.1 & 0.2 & 0.01 & $R_0 = 30\ h^{-1}$kpc, $R_\mathrm{dis}=R$ \\
P1B & {\bf 10} & bubble & 0.1 & 0.2 & 0.01 & $R_0 = 30\ h^{-1}$kpc, $R_\mathrm{dis}=R$ \\
P1C & {\bf 100} & bubble & 0.1 & 0.2 & 0.01 & $R_0 = 30\ h^{-1}$kpc, $R_\mathrm{dis}=R$ \\
&&\\[-3pt]
\multicolumn{7}{c}{Varying Mechanical Heating Efficiency}\\
\hline
P2A & 1 & bubble & 0.1 & {\bf 0.02} & 0.01 & $R_0 = 30\ h^{-1}$kpc, $R_\mathrm{dis}=R$ \\
P2B & 1 & bubble & 0.1 & {\bf 0.5} & 0.01 & $R_0 = 30\ h^{-1}$kpc, $R_\mathrm{dis}=R$ \\
&&\\[-3pt]
\multicolumn{7}{c}{Varying Feedback Frequency}\\
\hline
P3A & 1 & bubble & 0.1 & 0.2 & {\bf 0.001} & $R_0 = 30\ h^{-1}$kpc, $R_\mathrm{dis}=R$ \\
P3B & 1 & bubble & 0.1 & 0.2 & {\bf 0.1} & $R_0 = 30\ h^{-1}$kpc, $R_\mathrm{dis}=R$ \\
&&\\[-3pt]
\multicolumn{7}{c}{Varying Size of Feedback Region}\\
\hline
P4A & 1 & bubble & 0.1 & 0.2 & 0.01 & {\bf $\mathbf{R}_\mathbf{0} = \mathbf{15}\ \mathbf{h}^\mathbf{-1}$kpc, $\mathbf{R}_\mathrm{{\bf dis}}=\mathbf{R}$} \\
P4B & 1 & bubble & 0.1 & 0.2 & 0.01 & {\bf $\mathbf{R}_\mathbf{0} = \mathbf{5}\ \mathbf{h}^\mathbf{-1}$kpc, $\mathbf{R}_\mathrm{{\bf dis}}=\mathbf{R}$} \\
P4C & 1 & bubble & 0.1 & 0.2 & 0.01 & {\bf $\mathbf{R}_\mathbf{0} = \mathbf{15}\ \mathbf{h}^\mathbf{-1}$kpc, $\mathbf{R}_\mathrm{{\bf dis}}=\mathbf{0}$} \\
&&\\[-3pt]
\multicolumn{7}{c}{Varying Thermal to Kinetic Ratio}\\
\hline
P5A & 1 & jet & {\bf 0.02} & {\bf 0} & 0 & $r_\mathrm{ej}=2.5$\ kpc, $h_\mathrm{ej}=2.0$\ kpc \\
P5B & 1 & jet & {\bf 0.02} & {\bf 0.5} & 0 & $r_\mathrm{ej}=2.5$\ kpc, $h_\mathrm{ej}=2.0$\ kpc \\
P5C & 1 & jet & {\bf 0.02} & {\bf 1} & 0 & $r_\mathrm{ej}=2.5$\ kpc, $h_\mathrm{ej}=2.0$\ kpc \\ 
P5D & 1 & jet & {\bf 0.1} & {\bf 0} & 0 & $r_\mathrm{ej}=2.5$\ kpc, $h_\mathrm{ej}=2.0$\ kpc \\
P5E & 1 & {\bf bubble} & {\bf 0.1} & {\bf 0.2} & {\bf 0.0001} & {\bf $\mathbf{R} = \mathbf{2}\ \mathbf{h}^\mathbf{-1}$\ kpc, $\mathbf{R}_\mathrm{{\bf dis}}=\mathbf{0}$} \\
%&&\\[-3pt]
\hline
\hline
\end{tabular*}
\end{center}
\label{tbl:phys_par}
\end{table*}

We first explore `numerical' parameters that are required in the numerical implementations but are essentially arbitrary and often chosen based on numerical rather than physical considerations. Variations of these parameters are listed in Table \ref{tbl:num_bub_par} and Table \ref{tbl:num_jet_par} for the bubble and jet model, respectively. The tests are divided into groups, each of which examines one of the numerical parameters. We first do a convergence test by varying the peak resolution, $\Delta x$, for both the bubble model (group B1) and the jet model (group J1). Because the constant $\alpha$ model is motivated by the argument that the accretion rate is under-estimated due to resolution insufficient to resolve the Bondi radius, in group B2 we scale the $\alpha$ parameter with resolution to see if the results are consistent. In the jet model we do not vary $\alpha$ because we will show that the accretion rate is suppressed from the beginning and hence varying $\alpha$ has a minor effect. Since in the jet model the feedback is distributed in the very inner few kpc around the black hole, it is more likely to interfere with the peak resolution and the choice of the size of the region for gas accretion and depletion. Therefore, we experiment with scaling the jet size with resolution in group J2 and varying the accretion and depletion radii in group J3. The impacts of these numerical considerations will be discussed in \S~\ref{sec:numerical}. Note that for these jet runs the feedback and thermal efficiency parameters are chosen to be $\epsilon_\mathrm{f}=0.1$ and $\epsilon_\mathrm{m}=0\%$ (purely kinetic), as commonly adopted in previous jet models \citep{Gaspari, Dubois}.

Table \ref{tbl:phys_par} summarizes the model or parameter variations that are more physically motivated. Again they are separated into groups. The effects we investigate here include varying the accretion strength, the mechanical feedback efficiency, the threshold for triggering a feedback event, the size and center of the feedback region, and the ratio between thermal and kinetic energy. Here we note that despite the difference in the parameterizations used in the bubble and jet models, we do expect their results to overlap when bubbles are injected almost continuously (with a small $\delta_{\mathrm{BH}}$) into a small region centered on the black hole (run P5E), and when the jet is purely thermal (run P5C). The differences between these two runs would mainly be due to the shape of the injection region. We discuss the influence of these `physical' parameters in \S~\ref{sec:physical}.  

When not specifically mentioned, the parameters are fixed to their fiducial values: $\epsilon_{\mathrm{f}} = 0.1$, $\epsilon_{\mathrm{r}} = 0.05$, $\epsilon_{\mathrm{m}} = 0.2$, $R_0=30\ h^{-1}\ \mbox{kpc}$, $E_0=10^{55}\ \mbox{erg}$, $\rho_0=10^4\ h^2\ \mbox{M}_\odot\ \mbox{kpc}^{-3}$, $r_{\mathrm{ej}} = 3.2$\ kpc, and $h_{\mathrm{ej}} = 2.5$\ kpc. Note that these values are chosen to match those adopted in previous cosmological simulations \citep{Sijacki, Dubois}, which are able to successfully reproduce the observed black hole density, star formation history, and $L_X$--$T_X$ relation. Therefore it is also one of our aims to examine whether the same set of parameters could also recover the properties of the ICM within observational constraints. 

%===================================================

\section{Sensitivity Study}
\label{sec:sensitivity}

In this section we present a sensitivity study of all the relevant parameters in the AGN subgrid models. Since most of these parameters are not well constrained due to lack of knowledge of the detailed physical processes, we will first start from a relatively large parameter space and then examine whether the results are consistent with observed limits. By performing the sensitivity study we would like to identify those variables to which the ICM properties are most sensitive. At the same time, this study also provides us with information about which ICM properties are most robust to the uncertainties in the AGN subgrid models.    

\subsection{The Fiducial Run}
\label{sec:fiducial}

\begin{figure}%[tbp]
\begin{center}
\includegraphics[scale=0.45]{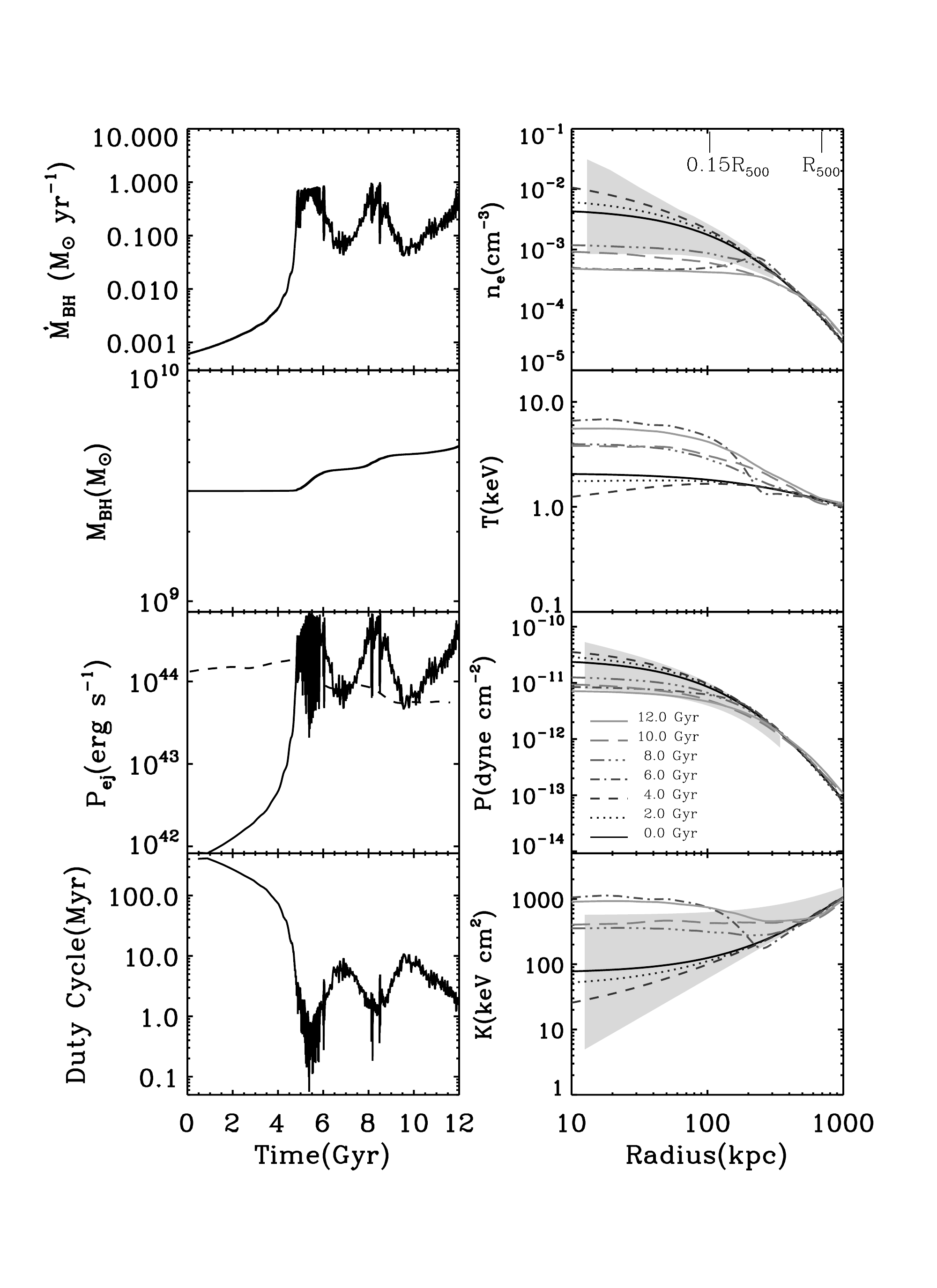} 
\caption{Results for the fiducial run P1A. {\it Left column (from top to bottom)}: Evolution of black hole accretion rate, black hole mass, power of feedback, and duty cycle. The X-ray luminosity inside the core ($R\leq 0.15R_{500}$) is overplotted with the injected power using the dashed line. {\it Right}: Radial profiles of gas density, temperature, pressure, and entropy. Grey areas are the observed ranges of density, pressure and entropy profiles from \citet{Croston}, \citet{Arnaud} and \citet{Cavagnolo}, respectively.}
\label{fig:fiducial}
\end{center}
\end{figure}

Here we show the general features of run P1A as an instructive example. This run uses the bubble model with parameters that are commonly adopted in previous cosmological simulations \citep{Sijacki}. The left column of Figure \ref{fig:fiducial} shows the evolution of the SMBH accretion rate, the mass of the black hole, the power of feedback (bubble energy divided by the duty cycle), and the duty cycle (time between two feedback events). The right column shows the profiles of gas density, temperature, pressure and entropy ($K\equiv T/n_e^{2/3}$) at different times. For this run, the accretion rate starts from a value of $\sim 6\times 10^{-4}\ \mbox{M}_\odot\ \mbox{yr}^{-1}$ ($\dot{M}_{\rm BH}/\dot{M}_{\rm Edd}\sim 9\times 10^{-6}$). The black hole grows slowly in its mass and generates weak bubbles only every few hundred Myr for the first $\sim 4$\ Gyr. The injected energy delays the strong cooling which would occur if there were no feedback. But it is still unable to balance radiative losses, so the accretion rate grows rapidly and reaches $1\%$ of the Eddington rate, triggering quasar-mode feedback at $t\simeq 5$\ Gyr. At this time the accretion rate is held near the value $1\ \mbox{M}_\odot\ \mbox{yr}^{-1}$ by powerful and frequent ($\sim 0.1-1$\ Myr) feedback events (both bubble and quasar modes), which in turn heat and expand the surrounding gas and cause the accretion rate to drop. The corresponding decrease of feedback then results in another round of strong cooling and feedback events. The cluster then fluctuates with a timescale of $\sim 3$\ Gyr till the end of the simulation. 

The gas profiles change according to the AGN activity. During the first $\sim 4$\ Gyr, the central density increases and the temperature decreases due to radiative cooling, similar to the case without AGN feedback. But once powerful feedback from the AGN starts to take place after $t\simeq 5$\ Gyr, the gas entropy within the injected bubbles is raised, so that the gas is heated and pushed outward. After later times the cluster oscillates around a quasi-static profile that is flatter and hotter than the initial profile. 

For the density, pressure and entropy profiles, we overplot our results with the observed profiles recently compiled for a large sample of clusters by \citet{Croston}, \citet{Arnaud} and \citet{Cavagnolo}, respectively. These observed profiles are remarkably uniform and self-similar at outer radii, while the dispersion increases toward the center. We note that despite these energetic AGN outbursts, the pressure profiles at all times lie well within the observed range. The universality of the pressure profiles is maintained because the density and temperature of the bubbles compensate each other to reach hydrostatic equilibrium with the surrounding ICM. 

On the other hand, the entropy profiles, though following a standard `power-law plus floor' profile in general, sometimes contrast with the observed profiles right after powerful AGN outbursts, e.g., the large floor and entropy inversions at $t=6$\ Gyr and $t=12$\ Gyr. These powerful events are also reflected in the density profiles, which have more flattened cores than observed. Note that the bubbles generated in this run have radii $\sim 100-200$\ kpc, larger than typical sizes of observed X-ray cavities \citep{McNamara}. Therefore, the influence of the bubbles may be overestimated due to the parameter choice of the bubble size (i.e.\ $R_0$ in Eq.\ \ref{eq:rbub}). As we will discuss in \S~\ref{sec:rbub}, though this problem can be alleviated by choosing a smaller $R_0$, it means that the default parameters for bubble sizes may not be applicable to every cluster but need to be fine-tuned. 

This run shows that such AGN feedback models can successfully self-regulate black hole growth, as also demonstrated in previous work \citep[e.g.][]{Sijacki}. It also produces cluster profiles that are in general consistent with observations. In the following subsections we will start varying the parameters to see how they affect the evolution of the AGN activities and cluster profiles. Note that our simulations do not include gas self-gravity. Though fragmentation of gas due to self-gravity could be important during the hierarchical formation of clusters \citep{Dubois12}, for the already-formed cluster in our simulations, the effect of gas clumping is minor because the Jeans length is large, except in the central regions where catastrophic cooling occurs for a few runs (see \S~\ref{sec:jet}). Also, including self-gravity would possibly deepen and flatten the cluster potential well and result in a flatter asymptotic pressure profile. However, since the main interest of this study is the differences due to model variations, and gas only contributes to a minor portion of the total cluster potential, we do not expect the inclusion of gas self-gravity to significantly change the results.

\subsection{Numerical Parameters}
\label{sec:numerical}

\begin{figure*}%[tbp]
\begin{center}
\includegraphics[scale=0.4]{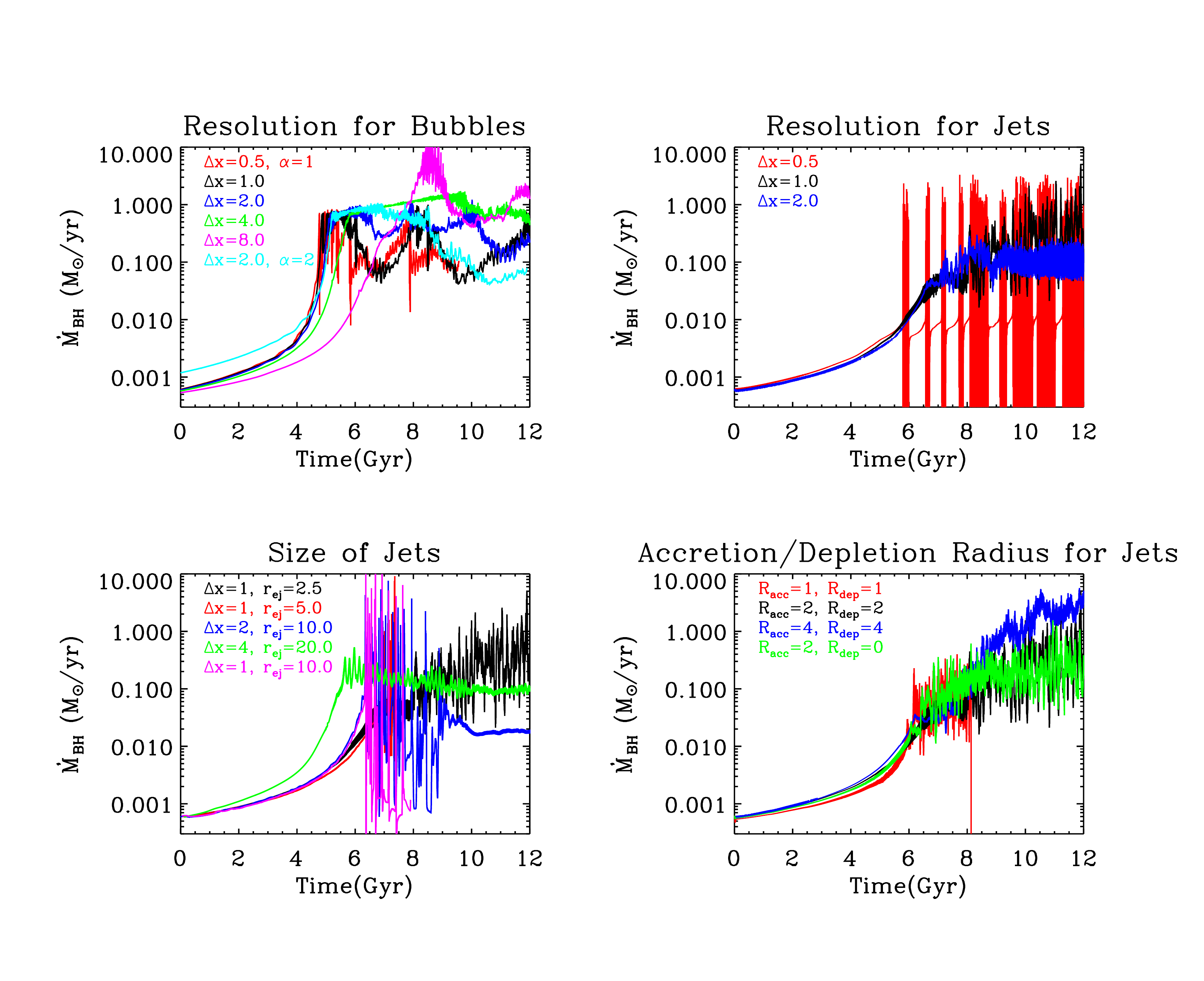} 
\caption{Black hole accretion rates for varying numerical parameters (see Table \ref{tbl:num_bub_par} and Table \ref{tbl:num_jet_par}). {\em Top left}: Varying the peak resolution (group B1) and scaled $\alpha$ with the resolution (group B2) for the bubble feedback model. {\em Top right}: Affecting the accretion rate by varying the peak resolution in the jet feedback model (group J1). {\em Bottom left}: Effects of different jet sizes with fixed/scaled peak resolution (group J2). {\em Bottom right}: Results of changing the radius for computing the accretion rate and the radius for removing the accreted gas in units of zones (group J3).}
\label{fig:num}
\end{center}
\end{figure*}

In this section we present results for varying the numerical parameters both in the bubble model (Table \ref{tbl:num_bub_par}) and the jet model (Table \ref{tbl:num_jet_par}). For the bubble model, we study the effect of peak resolution, as well as scaling the $\alpha$ parameter with resolution. For the jet model, we test variations of the resolution, jet sizes, and the radius for computing accretion rate and for removing gas. The definitions and explanations for these parameters can be found in \S~\ref{sec:models}. We probe the sensitivity to these parameters by examining the evolution of the black hole accretion rate. The cluster profiles are not plotted here since they are closely related to the accretion rate, as seen in the previous section.

The convergence test for the bubble model is shown in Figure \ref{fig:num} (top left). We allow a large range of variation for the peak resolutions (as large as 8\ kpc) because current high-resolution cosmological simulations can typically reach resolutions of a few kpc, but to go beyond that is progressively more difficult due to computational costs. Therefore it is important to understand whether such simulations with subgrid AGN models are numerically converged. We find that as the peak resolution is increased, fluctuations with shorter timescales and with larger amplitudes are captured, whereas for runs with degraded resolutions the accretion rates react more slowly, reach the first peak later, and have less variation. This makes the SMBH in the lower-resolution runs increase its mass at a later time, but it grows to a larger value at the end of the simulation by time 12\ Gyr. In particular, the final black hole mass for run B1E ($\Delta x=8$\ kpc) differs from run B1B ($\Delta x=1$\ kpc) by about a factor of 2. Therefore simulations with poorer resolutions may underestimate the variations in accretion rate and corresponding ICM properties. They may also find more massive black hole populations when other conditions are held the same.

Since the constant $\alpha$ in the bubble model is often invoked to compensate for the underestimation of the accretion rate due to resolution, in principle a larger $\alpha$ should be used when the resolution is poorer. In group B2 we test whether such scaling of $\alpha$ would help account for the difference in resolution. As shown in Figure \ref{fig:num} (top left), the initial accretion rate for run B2A ($\Delta x=2$\ kpc, $\alpha=2$) is boosted compared to the run with the same peak resolution (B1C). However, since this cluster has a flat initial entropy profile (recall that $\dot{M}_{\rm Bondi} \propto n_{\rm e}/c_{\rm s}^3 \propto n_{\rm e}/T^{3/2} \propto K^{-3/2}$), this boost is not necessary to match the accretion rates of the higher-resolution runs (B1A and B1B).
%this boost does not help make the accretion rate approach the values of the higher-resolution runs (B1A and B1B), simply because this cluster has a flat initial entropy profile. 
This points out one problem with the $\alpha$ accretion model, which is that a single constant value of $\alpha$ may not be appropriate for a population of clusters with various core profiles. At later times when cooling and feedback events get stronger, the evolution becomes nonlinear and depends sensitively upon the detailed interactions between the bubbles and quasars with the surroundings. So the outcome of run B2A is close to neither run B1C nor run B1B. Therefore, simply scaling $\alpha$ by a constant factor does not in general work to compensate for the change in resolution. In fact, we will see in \S~\ref{sec:accretion} that choosing the value of $\alpha$ is nontrivial and has a great impact on the evolution of SMBH and ICM properties. 

For the jet feedback model, we also find more variation in the accretion rate for runs with higher peak resolution (Figure \ref{fig:num}, top right). Despite the difference in the amplitude of fluctuations, the mean accretion rates are more robust to the resolution than in the bubble model, at least when the zone size is larger than 1\ kpc (run J1B and J1C). 

The dramatic change in the accretion rate for run J1A may be understood in combination with the results of scaling jet sizes with resolution (Figure \ref{fig:num}, bottom left). This problem of large amplitude fluctuations occurs in all the runs where the jet size is much greater than the size of a grid cell (J2A-J2D). Recall that the radius for computing the accretion rate is set to 2 zones. Therefore when the accretion radius is small compared to the region used to apply jet feedback, such accretion rate estimates are sensitive more to the details in the feedback itself than the actual accretion from the ICM. The only exception is run J2C, which has very extended jets that have essentially the same effects as the bubble feedback. In particular, this run produces results that are comparable to bubbles with high frequencies (run P3A) and small sizes (run P4C) that will be shown in later sections. 

Finally we vary the radius for computing the accretion rate, $R_\mathrm{acc}$, and the radius for removing the accreted gas, $R_\mathrm{dep}$ (Figure \ref{fig:num}, bottom right). We find that either changing the accretion and depletion radii, or removing gas depletion altogether, has a minor effect on the results. Note however that run J3A crashed at $t \simeq 8$\ Gyr with a sudden drop in the accretion rate. Since in this run the accretion radius is only one zone and is smaller than the jet size, the accretion rate is very sensitive to the central few zones within the feedback region. A slight displacement of gas within the innermost cells causes a sudden reduction in the accretion rate and the associated feedback, which induces an unphysical surge of gas into the central zones. Therefore, for numerical stability we recommend using an accretion radius larger than the region of jet feedback.

To summarize briefly, we find that increasing the peak resolution generally produces more variable accretion rates. Bubble feedback suffers a greater influence when varying the resolution, whereas jet feedback is more robust, as long as the accretion radius is carefully chosen (larger than the jet size). 

\subsection{Physical Parameters}
\label{sec:physical}

\subsubsection{Dependence on Accretion Models}
\label{sec:accretion}

\begin{figure*}%[pt]
\begin{center}
\includegraphics[scale=0.45]{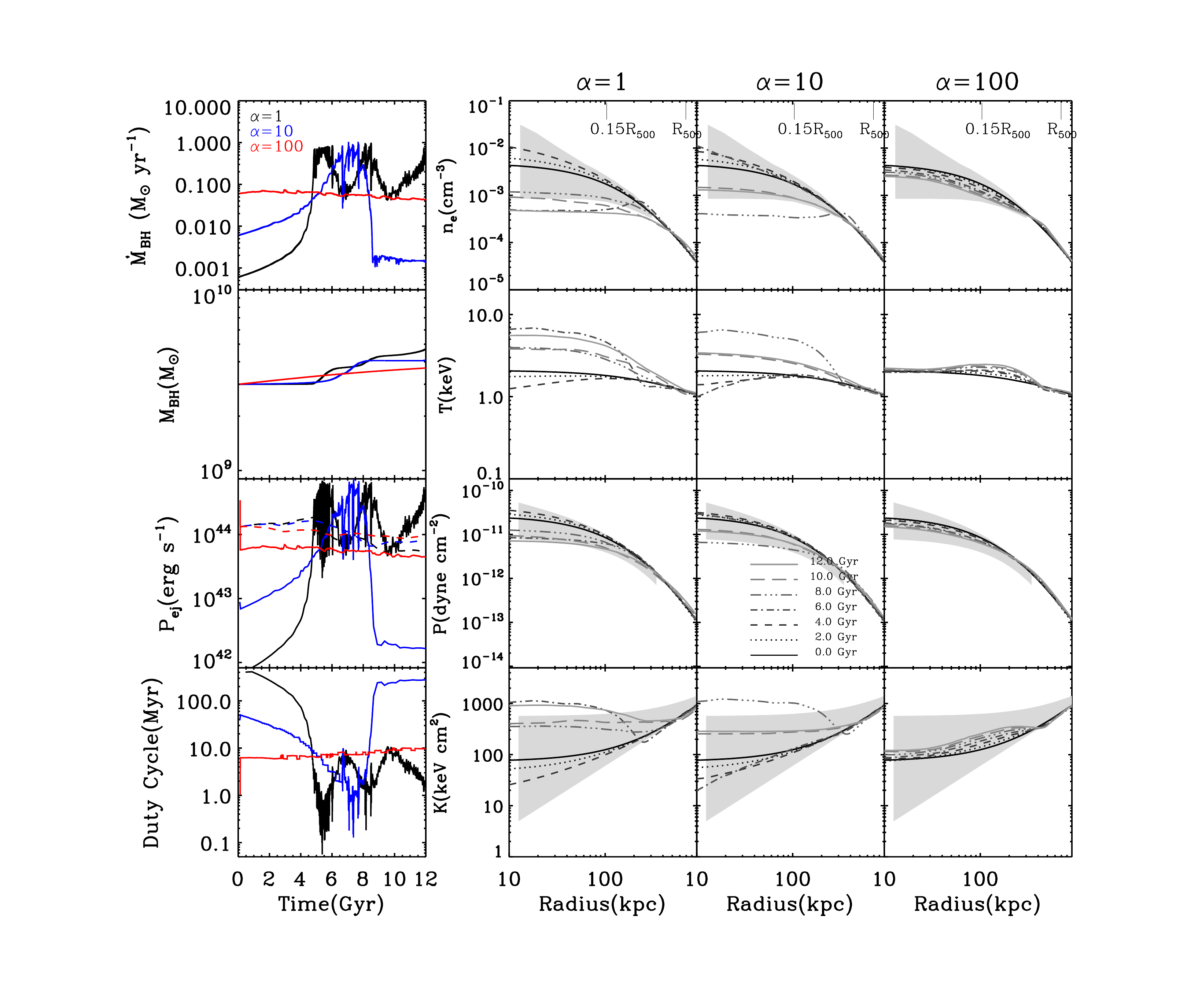} 
\caption{Evolution of AGN activity and cluster profiles for different accretion models (group P1). Symbols are the same as in Fig \ref{fig:fiducial}.}
\label{fig:accretion}
\end{center}
\end{figure*}

Table \ref{tbl:phys_par} lists the variations of the physically-motivated parameters under consideration. For the first group of runs we vary the method of computing the accretion rate. %Note that the $\beta$ model is the same as $\alpha=1$ in the simulations using the bubble model since the feedback is effective in suppressing the central gas density below $0.1\ \mbox{cm}^{-3}$. Since the cluster has a flat entropy profile at the center, $\alpha=1$ is justified based on the resolution argument for the Bondi accretion. However, in reality the accretion rate is still likely to be underestimated because of the unresolved multiphase gas. For example, the observed Bondi accretion rate for the SMBH in M87 is estimated to be $0.026\ \mbox{M}_\odot\ \mbox{yr}^{-1}$ at its Bondi radius \citep{Allen}, which is about 30 times higher than the rate computed from the central entropy of our simulated cluster. Therefore, by choosing $\alpha$ values ranging from 1 to 100, we expect to cover situations where the actual accretion rate is underestimated as well as where it is overestimated. 
As can be seen from Figure \ref{fig:accretion}, the evolution of the cluster is very different for different values of the accretion strength parameter, $\alpha$. For run P1A where $\alpha=1$, the accretion starts from a small value and thus the initial feedback power is small compared to the core X-ray luminosity. Here strong cooling occurs and triggers cycles of feedback events at later times, as described in \S~\ref{sec:fiducial}. On the other hand, in run P1C ($\alpha=100$), the feedback power in the beginning is already comparable to the X-ray luminosity, so the cluster never goes through the strong cooling phase and is roughly in equilibrium throughout the simulation time. The cluster profiles respond to the AGN activity in the same way as in the fiducial run, which is the reason why the gas properties have more fluctuations in run P1A than P1C.  

The evolution of the cluster core can be further quantified using the entropy floor \citep{Cavagnolo}, slopes of the density profile \citep{Croston} and entropy profile \citep{Sanderson}, or the cooling time \citep{Mittal}. Following the definition in \citet{Mittal}, we categorize cluster cores as strong cool cores (SCC; $t_{\rm cool}<1$\ Gyr), weak cool cores (WCC; $1\ \mbox{Gyr}<t_{cool}<7.7$\ Gyr), and non cool cores (NCC; $t_{\rm cool}>7.7$\ Gyr). Based on this definition, the cluster starts with a WCC with $t_{\rm cool}\sim 7.5$\ Gyr. In all the bubble models studied here, the cluster never reaches the SCC state because the overall heating is very effective. For run P1A, the cooling time drops to its minimum of $t_{\rm cool}\sim 2$\ Gyr at $t\simeq 4$\ Gyr and climbs up to a NCC in the end. Run P1B is similar to P1A. For run P1C, the cooling time instead keeps rising, so the cluster became a NCC cluster soon after the start of the simulation.  

The large influence of the assumed accretion model on the evolution history of the cluster poses a great concern for simulations including AGN feedback. It implies that given the uncertainties in the accretion mechanisms, current AGN subgrid models have very limited power to predict the evolution of cluster core properties. That is, simulations with different accretion models can produce very different results, e.g., the fraction of CC versus NCC clusters as a function of time. Note that the cosmological simulation of \citet{Sijacki} uses $\alpha=100$, which is very effective in suppressing the formation of cool cores as we have shown. Indeed their simulations generally overpredict the fraction of NCC clusters at the present day compared to observations.\footnote{private communication} Therefore, in order to produce robust results, accurate modeling of the accretion onto the SMBH is crucial. 

\subsubsection{Varying the Mechanical Heating Efficiency}
\label{sec:effm}

\begin{figure*}%[pt]
\begin{center}
\includegraphics[scale=0.45]{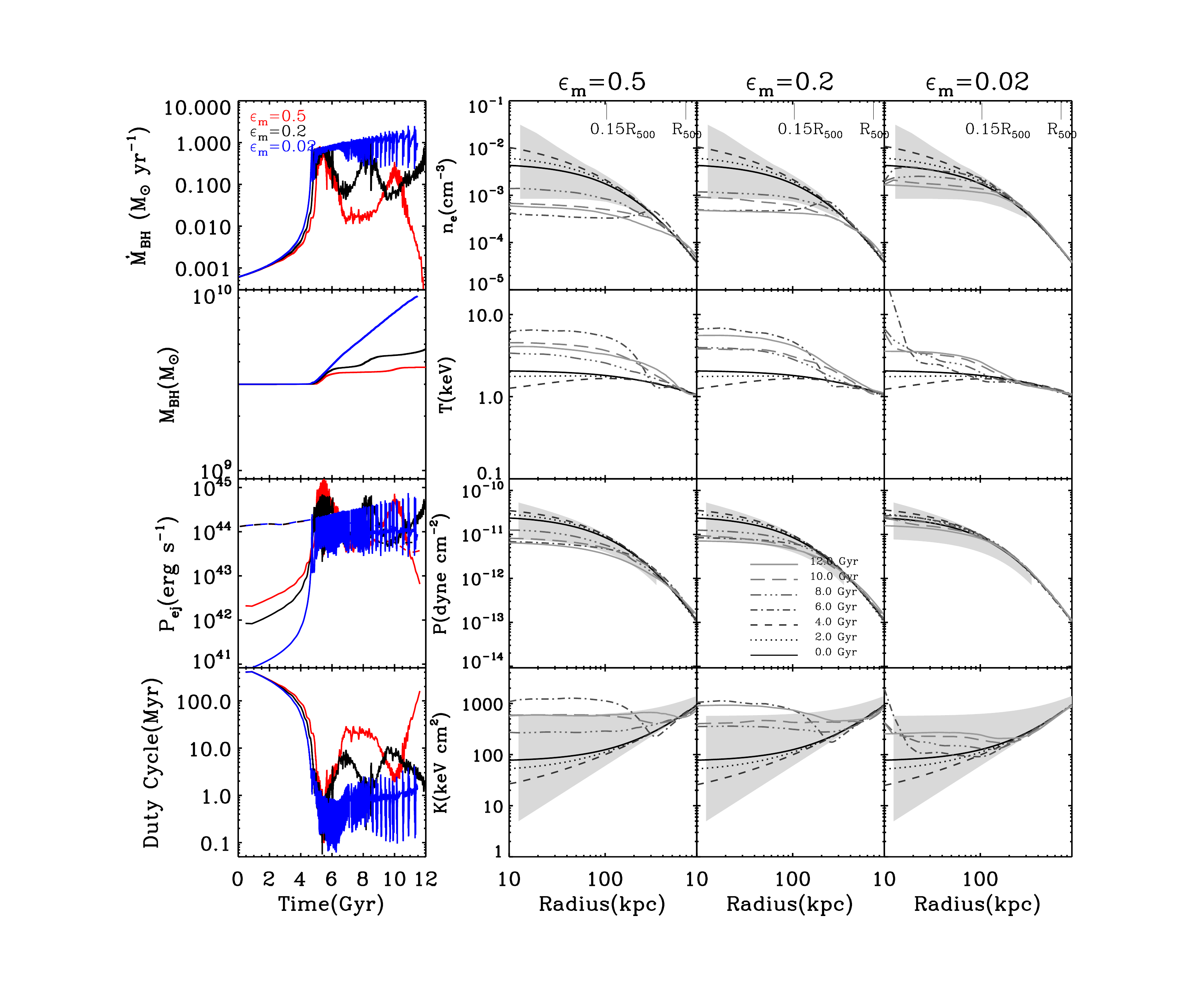} 
\caption{Evolution of AGN activity and cluster profiles for different mechanical heating efficiency $\epsilon_\mathrm{m}$ (group P2 plus the fiducial run P1A).}
\label{fig:effm}
\end{center}
\end{figure*}

The mechanical heating efficiency, $\epsilon_\mathrm{m}$, parametrizes how much feedback energy is actually converted into thermal energy and used as a source of heating. Observationally the ratio of cavity power to the Bondi accretion rate is estimated to be a few percent \citep{Allen}, which motivates previous simulations to adopt similar values for the net efficiency, $\epsilon_\mathrm{f} \epsilon_\mathrm{m}$. However, it is still unclear how the cavity power is converted into heat and how long this process takes. If the bubbles do not mix with the ICM efficiently, in principle the mechanical heating efficiency could be much lower. For example, \citet{Vernaleo} used purely hydrodynamic simulations and estimated the fraction of injected kinetic energy going into internal energy (i.e.\ $\epsilon_{\rm m}$) to be only a few percent. But since the actual magnitude of mechanical heating would depend on the details of mixing, simulations with more realistic physical treatment of the ICM are required to pin down this number. Here we probe the range $\epsilon_\mathrm{m}=0.02-0.5$, which is permitted by current constraints and covers values commonly used in previous simulations.    

Figure \ref{fig:effm} shows the SMBH and cluster evolution with varying $\epsilon_\mathrm{m}$ (group P2). In general the changes in $\epsilon_\mathrm{m}$ do not alter the fate of the cluster, in contrast to the variation in the accretion models discussed in the previous section. For all three runs, the cluster goes through gradual cooling for the first 4\ Gyr, which eventually grows and triggers a sequence of feedback events, just like the fiducial run P1A. The more powerful bubbles for run P2A ($\epsilon_\mathrm{m}=0.5$) only delay the time of strong cooling a little bit, but for the first 4\ Gyr the results are almost indistinguishable. After 5\ Gyr, the cluster again starts to oscillate among states that are closely related to the feedback activity. When the mechanical heating efficiency is larger, the feedback is less frequent, more powerful and more effective in reducing the accretion rate. The reduced accretion rate takes a longer time to grow back to 1\% of the Eddington rate, so the cluster fluctuates with a timescale of $\sim 5$\ Gyr for run P2A ($\epsilon_\mathrm{m}=0.5$), while for run P2B ($\epsilon_\mathrm{m}=0.02$) the cluster oscillates with smaller amplitude and timescale.    

\begin{figure*}%[ht]
\begin{center}
\includegraphics[scale=0.45]{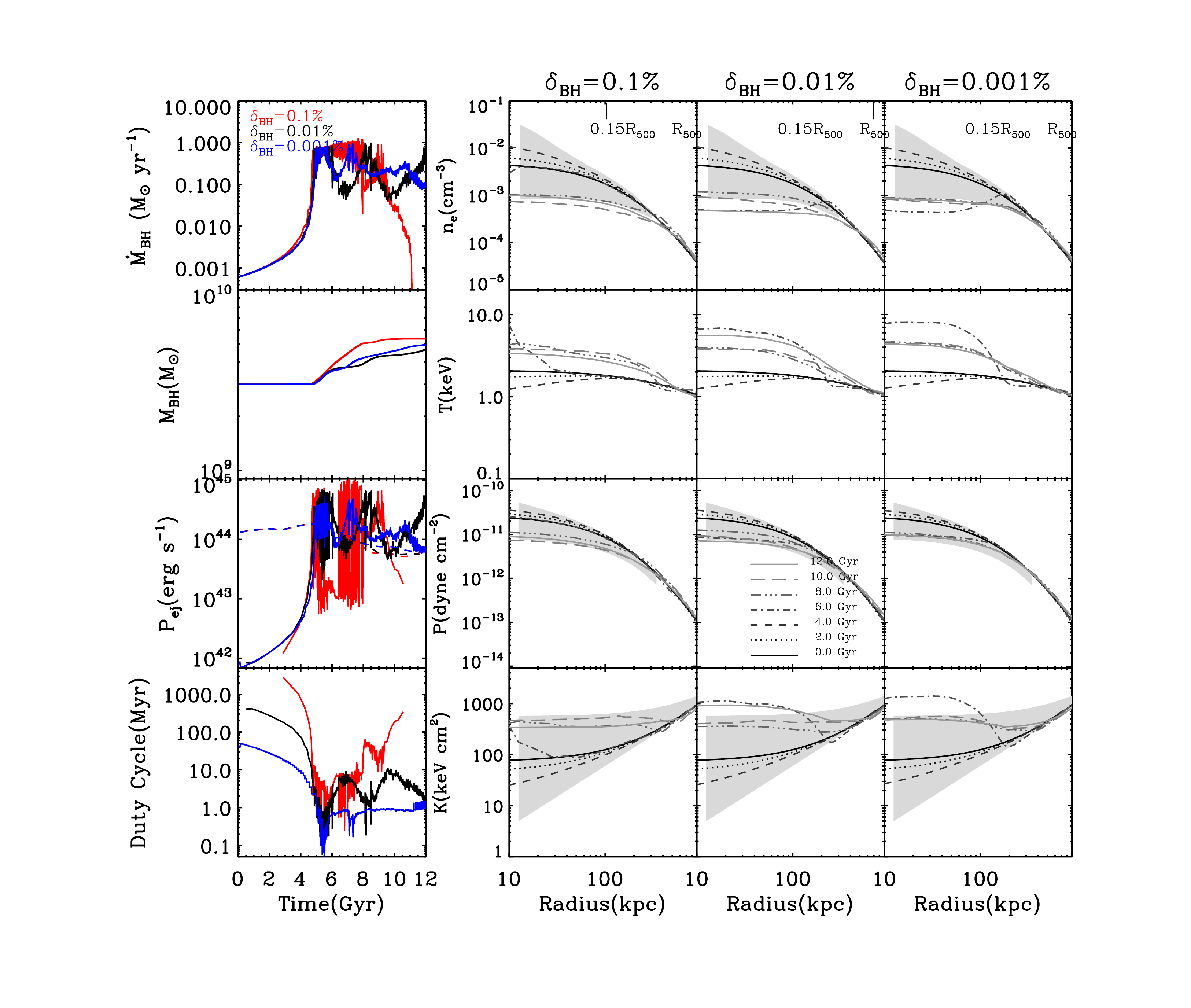} 
\caption{Dependence on the threshold for bubble injection (group P3 plus the fiducial run P1A). The threshold, $\delta_\mathrm{BH}$, is defined to be the minimal fractional increase in black hole mass required to inflate a bubble. Smaller thresholds generate more frequent bubbles.}
\label{fig:dbh}
\end{center}
\end{figure*}

Despite the overall similarity of behavior, the change in $\epsilon_\mathrm{m}$ does result in several noticeable trends. First, the growth of the black hole mass is sensitive to the efficiency. For larger efficiencies, the accretion rate is suppressed so that the black hole does not grow as much as when the efficiencies are smaller. Second, efficient mechanical heating produces large density plateaus and entropy floors in the cores. Particularly at moments right after an energetic outburst, the density and entropy profiles can have temporary excursions that go beyond observed ranges. This problem may be alleviated by reducing the scaling of the bubble radius (see \S~\ref{sec:rbub}). Lastly, for runs with small efficiencies, the temperature profiles sometimes have a peak at the center. This is due to the concentrated thermal energy injection in the quasar mode, which is invoked more frequently in these cases to help the bubbles at high accretion rates. This is an issue shared with all the models which input thermal energy into a small region, such as those jet models with nonzero thermal components (see \S~\ref{sec:jet}). To avoid this undesirable feature of too much quasar-mode feedback, we did a test run identical to P2B ($\epsilon_\mathrm{m}=0.02$), but with the switch to quasar mode turned off (i.e., the bubble mode is applied throughout the simulation). %with quasar mode turned off. 
We find that pure bubble heating alone with such small efficiencies is insufficient to halt the cooling catastrophe. The accretion rate increases to $\sim 100\ \mbox{M}_\odot\ \mbox{yr}^{-1}$ at $t\simeq 5$\ Gyr and generates a huge bubble that essentially blows all the cluster gas away. This puts a lower limit on the mechanical feedback efficiency, as also found by \citet{Gaspari11}. 

\subsubsection{Varying the Feedback Frequency}
\label{sec:dbh}

In the bubble model employed in this paper \citep{Sijacki}, the bubbles are inflated when the black hole mass increases its mass by a fraction $\delta_\mathrm{BH}$. Therefore, a larger $\delta_\mathrm{BH}$ corresponds to a longer duration between successive bubble events, as can be seen from Figure \ref{fig:dbh}. The exact relationship between the parameter $\delta_\mathrm{BH}$ and the duty cycle actually can be derived given the criterion for triggering bubbles. For this particular bubble model, the increase in BH mass between subsequent AGN outbursts is $\delta_{\rm BH} M_{\rm BH} \simeq \dot{M}_{\rm BH} \Delta \tau$. Therefore, 
\begin{equation}
\Delta \tau \simeq
10^3\ \delta_{\rm BH} \left(\frac{M_{\rm BH}}{10^9\ \mbox{M}_\odot} \right)
\left( \frac{\dot{M}_{\rm BH}}{1\ \mbox{M}_\odot\ \mbox{yr}^{-1}} \right)^{-1} \ \mbox{Myr}.
\label{eq:tau}
\end{equation}

\begin{figure*}%[th]
\begin{center}
\includegraphics[scale=0.45]{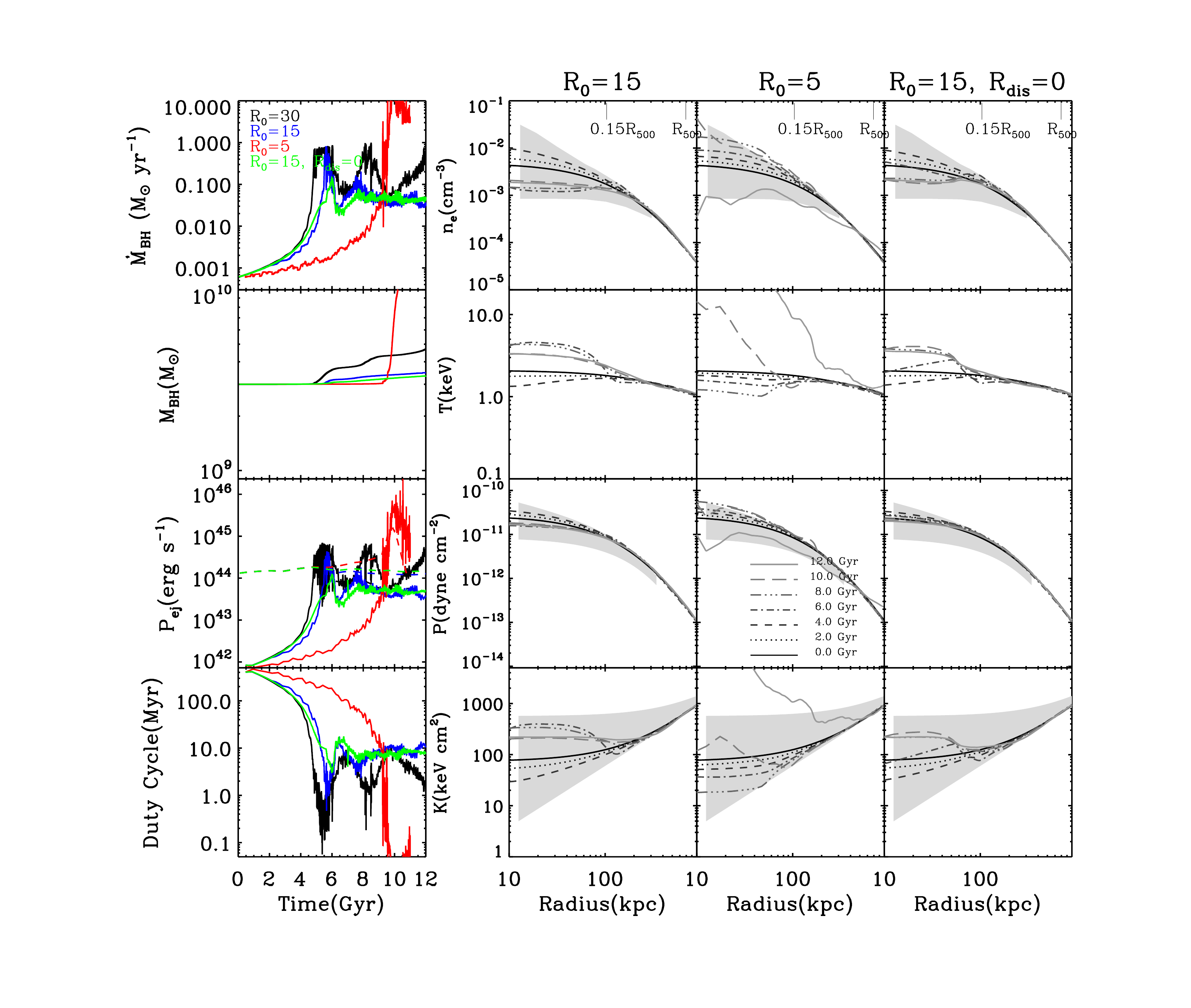} 
\caption{Effect of the region for bubble injection (group P4). $R_0$ is the scaling coefficient for bubble radii as in Eq.\ \ref{eq:rbub}. $R_{\rm dis}$ is the displacement of bubble center from the AGN. The evolution of the SMBH for the fiducial case P1A is plotted using black curves, but its cluster profiles are omitted here (see previous figures).}
\label{fig:rbub}
\end{center}
\end{figure*}

Indeed, we can see from Figure \ref{fig:dbh} that the initial duty cycle scales with $\delta_{\rm BH}$, and that the evolution of the duty cycle is roughly inversely proportional to the accretion rate. For M87, if we take observationally constrained values, $\dot{M}_{\rm BH}\sim 0.026\  \mbox{M}_\odot\ \mbox{yr}^{-1}$, $M_{\rm BH}\sim 3\times 10^9\ \mbox{M}_\odot$ \citep{Allen}, and $\Delta \tau \sim 10^7$ yr \citep[e.g.][]{Million10}, we obtain $\delta_{\rm BH}\sim 0.01\%$, which is consistent with the value used in the fiducial run. 
%\sim 10^{-4}$. Thus for the fiducial model ($\alpha=1$), $\delta_{\rm BH}\sim 0.01\%$ is consistent with observations. 
Note that Eq.\ \ref{eq:tau} is only true for this specific bubble model. The duty cycle would have different scalings with BH mass and accretion rate when a  different criterion is used to trigger bubbles. For example, in the model of \citet{Booth}, the injected energy per bubble is fixed according to a minimum heating temperature, so that $E={\rm constant}\propto \dot{M}_{\rm BH} c^2 \Delta \tau$ gives $\Delta \tau \propto \dot{M}_{\rm BH}^{-1}$ with no explicit dependence on BH mass. On the other hand, the model of \citet{Battaglia} sets $\Delta \tau={\rm constant}$ instead. Therefore, how to trigger bubbles in the AGN models is not completely arbitrary, but in principle can be constrained by observed scalings of duty cycles \citep[e.g.][]{Shabala08}, though these measurements are themselves very difficult.       

Fortunately, we find that the frequency of injections in the model is not critical for the results. As shown in Figure \ref{fig:dbh}, the amplitudes of variation in the accretion rate and gas properties are similar for different $\delta_\mathrm{BH}$. The increase in temperature and entropy is slightly higher for smaller $\delta_\mathrm{BH}$, mainly because the injected energy is distributed in a smaller region, as the bubble sizes scale with the injected energy, which is smaller for smaller $\delta_\mathrm{BH}$ (Eq.\ \ref{eq:ebub}). But compared with the effects of accretion and mechanical heating efficiency, varying $\delta_\mathrm{BH}$, or the frequency of bubble injection, does not have as large an impact on the overall evolution of SMBH and the ICM. 

\subsubsection{Dependence on the Region of Feedback}
\label{sec:rbub}

Here we explore the effect of varying the region of bubble injection (group P4), including the scaling coefficient for bubble radii, $R_0$, as well as the displacement from the central AGN, $R_\mathrm{dis}$. As in the fiducial case, run P4A and P4B inject bubbles whose centers are randomly displaced within a sphere of radius $R_\mathrm{dis}$, but with smaller bubble radii. The typical size of bubbles is 100-200\ kpc, 50-100\ kpc and 20-30\ kpc for the fiducial run P1A, run P4A and run P4B, respectively. As shown in Figure \ref{fig:rbub}, reducing the bubble sizes is very effective in suppressing the accretion rate because of concentrated heating. The influence on the evolution of the cluster core is even more than increasing the heating efficiency (Figure \ref{fig:effm}). In other words, the evolution of SMBH and ICM properties is very sensitive to the choice of bubble radii.  

\begin{figure*}%[th]
\begin{center}
\includegraphics[scale=0.45]{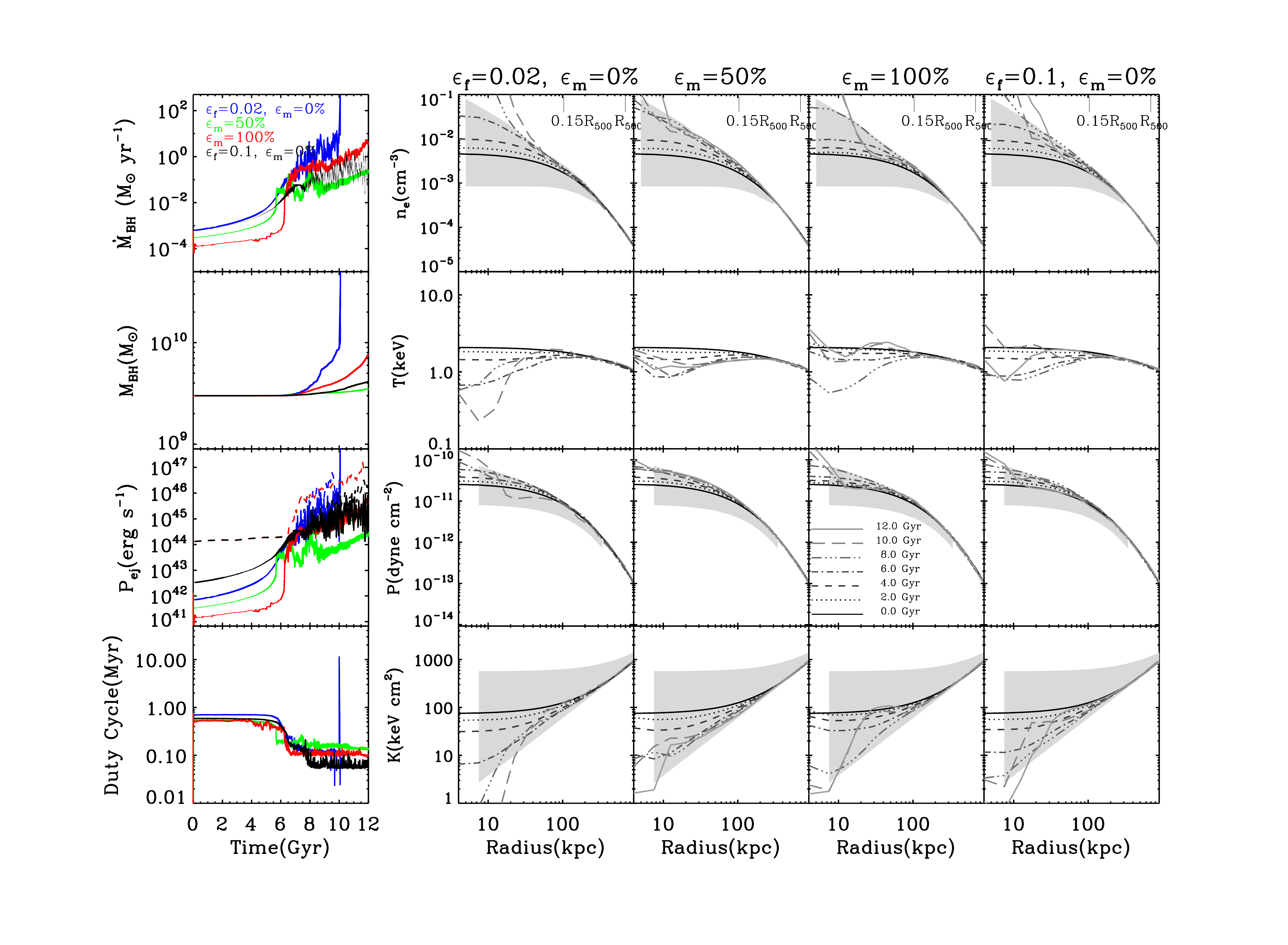} 
\caption{Varying the thermal and kinetic efficiencies in the jet feedback model (group P5A-P5D). $\epsilon_\mathrm{f}$ is the feedback efficiency, i.e., the ratio between total injected energy and the rest mass energy of the SMBH, and $\epsilon_\mathrm{m}$ is the fraction that goes into thermal energy.}
\label{fig:jets}
\end{center}
\end{figure*}

\begin{figure}
\begin{center}
\includegraphics[scale=0.37]{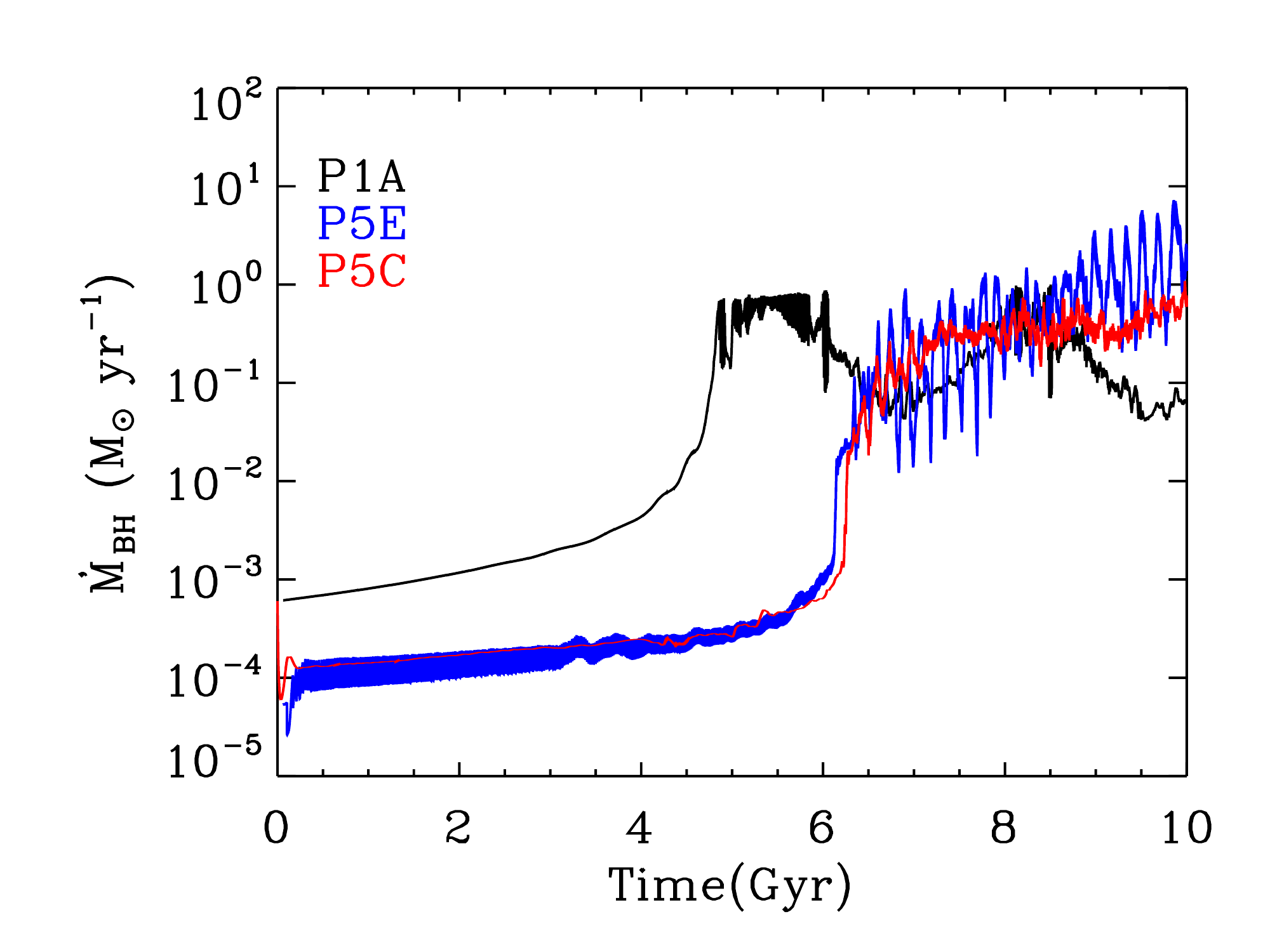} 
\caption{Comparing the bubble and jet feedback models. Run P1A is the fiducial bubble case, which generates large, randomly-displaced bubbles. Run P5C and P5E have the same net feedback efficiencies, $\epsilon_\mathrm{f}\epsilon_\mathrm{m}=0.02$. But run P5C uses purely thermal jets, while run P5E uses tiny, fixed, and almost continuous bubbles to mimic the jet run P5C. This figure demonstrates that the bubble and jet models are numerically consistent when appropriate parameters are chosen. Moreover, their differences are mainly caused by the size of the feedback region.}
\label{fig:bub+jet}
\end{center}
\end{figure}

Compared with the fiducial run, the cluster profiles for run P4A ($R_0=15\ h^{-1}$~kpc) have a smaller flattened core since the energy injection is more concentrated. Note that now the density and entropy profiles at all times are consistent with the observed range. This implies that the choice of bubble radii is not completely arbitrary. Bubbles that are too large may produce density and entropy profiles that are inconsistent with observations.

The sizes of the bubbles cannot be too small either. For run P4B ($R_0=5\ h^{-1}$~kpc) where small bubbles are randomly injected around the black hole, though concentrated heating greatly slows down cooling and accretion, catastrophic cooling still occurs at $t\simeq 10$\ Gyr and generates bubbles that dramatically heat the core. This may be due to the fact that the radius within which the cooling time is less than 10\ Gyr is around 100\ kpc. So the bubbles only heat a small fraction of gas within the cooling radius, while a substantial amount of gas is still allowed to cool and flow to the center at later times. 

We also did one run with the center of bubbles fixed on the central AGN (run P4C, $R_\mathrm{dis}=0$). Other parameters are the same as in run P4A ($R_0=15\ h^{-1}$kpc, $R_\mathrm{dis}=R$). From Figure \ref{fig:rbub} we can see that these two runs produce almost identical results, except that fixed bubbles tend to produce less smooth profiles than randomly-positioned bubbles. But these differences are minor. Therefore, as long as the size of bubbles and thus the input energy density are comparable, where around the SMBH to dump the energy has a lesser effect.

In summary, the size of the region used to inject thermal energy (but not so much the displacement from the black hole) is crucial in predicting the evolution of the SMBH and the ICM. Bubbles that are too large push too much gas outward and raise the entropy floor to an unrealistic level, while bubbles that are too small may not be able to heat all the region where it is needed, allowing catastrophic cooling. Therefore, for the current bubble model, or any model that requires setting the size of energy injection by hand, it would be difficult to find one parameter or scaling that works for all clusters. And even if the results are permitted by observed limits, the predictions would still be very sensitive to the chosen bubble sizes. 

\subsubsection{Effect of Thermal to Kinetic Ratio}
\label{sec:jet}

In this section we explore the models where the feedback energy is discharged in the form of jet, which has a shape function (Eq.\ \ref{eq:psi}) aligned with the z-axis, as opposed to the spherical bubbles discussed in previous sections. In our generalized parametrization of the jets (Eq.\ \ref{eq:jets}), the amount of thermal energy and kinetic energy can be tuned using two parameters: $\epsilon_\mathrm{f}$, the feedback efficiency, or the ratio between total injected energy and the rest mass energy of the SMBH, and $\epsilon_\mathrm{m}$, the fraction that goes into thermal energy. The comparison of different thermal to kinetic ratios (group P5A-P5D) is displayed in Figure \ref{fig:jets}.

The first thing to note for the jet models is that the accretion rates are immediately reduced to $\sim 10^{-4}\ \mbox{M}_\odot\ \mbox{yr}^{-1}$ as soon as the simulation starts and are more so when the thermal efficiency $\epsilon_\mathrm{m}$ is greater. This suppression is due to the fact that in the jet models, energy is injected within only a few kpc around the AGN, instead of large bubbles that extend up to tens or hundreds of kpc. As shown in the previous section, decreasing the size of the region used to distribute thermal energy can suppress the accretion rate very effectively. Therefore, the jets just resemble tiny bubbles. In order to verify whether the bubble and jet models are consistent under similar conditions, we did a run where bubbles with fixed radii $2\ h^{-1}$\ kpc are generated almost continuously (run P5E). Figure \ref{fig:bub+jet} shows that this run indeed reproduces the case P5C of purely thermal jets of identical net feedback efficiencies, $\epsilon_\mathrm{f}\epsilon_\mathrm{m}=0.02$. The small fluctuations for run P5E just reflect the fact that bubbles are produced every few timesteps rather than perfectly continuously. This test demonstrates that the bubble and jet models are numerically consistent and are degenerate when appropriate parameters are chosen. Moreover, it again emphasizes the point that the choice of the size for energy injection in the AGN subgrid models is nontrivial.

Runs P5A-P5C compare different ratios of thermal to kinetic energy, with the total feedback efficiency $\epsilon_\mathrm{f}$ kept fixed at 0.02. The fraction of energy that goes into thermal energy is 0\%, 50\%, and 100\% for run P5A, P5B, and P5C, respectively. As shown in Figure \ref{fig:jets}, all these jet models in general produce similar overall evolution of the accretion rates and cluster profiles. The accretion is halted by concentrated feedback from the beginning so the black hole grows very slowly for the first few Gyr. Since the injected power is smaller than the X-ray luminosity, the cluster gas cools more and more rapidly to a cool-core state at $t \simeq 6$\ Gyr, The rapid cooling quickly feeds the black hole and increases its mass, which allows the jets to stabilize cooling afterwards. The only exception is run P5A, which fails to overcome catastrophic cooling at $t \simeq 10$\ Gyr. So self-regulation of black hole growth may not be achieved by purely kinetic feedback with efficiencies that are too small.

As expected, the accretion rate is initially more suppressed for higher thermal efficiencies. But interestingly, after $t \simeq 6$\ Gyr run P5B ($\epsilon_\mathrm{m}=50\%$) becomes the most effective.            In other words, the most effective way to stifle cooling is not necessarily to dump all the feedback energy in thermal form, but rather a combination of thermal and kinetic feedback that facilitates mixing of the heated gas with the surroundings.  

We also performed another run P5D with purely kinetic feedback but higher total feedback efficiency, $\epsilon_\mathrm{f}=0.1$. Like run P5A, which is also kinetic but with $\epsilon_\mathrm{f}=0.02$, the initial accretion rate is not affected in the beginning since the kinetic energy has not transformed into heat. Their differences become more evident as feedback energy is thermalized and as the jets become more powerful after $t \simeq 6$\ Gyr. Note that at later times, the level of suppression is comparable to run P5B. So both raising the total feedback efficiency and tuning the thermal to kinetic ratio can slow down the accretion and black hole growth.  

It is also instructive to compare run P5D ($\epsilon_\mathrm{f}=0.1$, purely kinetic) with run P5C ($\epsilon_\mathrm{f}=0.02$, purely thermal). If their results are comparable, it would imply that 20\% of the kinetic energy is converted into heat, or a mechanical heating efficiency of 0.2, which is around values commonly assumed in AGN subgrid models using purely thermal feedback \citep{Sijacki, Booth}. Recent cosmological simulations also have found that either using purely kinetic feedback or purely thermal feedback assuming 15\% for the mechanical heating could match the local black hole properties \citep{Dubois11}. Here we find that run P5D has somewhat better ability than run P5C to halt cooling, implying possibly a mechanical heating efficiency higher than 20\%. Note that small discrepancies are expected because of different sizes of thermal feedback used by different simulations. For our jets the feedback region is confined within the small shape function, whereas simulations mentioned above used either extended bubbles \citep{Sijacki} or the nearest SPH particle \citep{Booth, Dubois11}. Note also that these arguments are based on simple hydrodynamic simulations, and would change if gas mixing is modified by additional physical effects, such as subgrid turbulence, viscosity, etc.

The resulting cluster profiles again reflect the ability of jets to stop cooling. Since run P5B is the most effective, its temperature decreases and density increases the least. However, in general the jets do not expel the gas or heat the gas as much as the bubbles, maintaining the core in the CC state. Note that the density profiles show a dense core of size $\sim 10$\ kpc at later times. This is due to a torus of cold gas that forms around the black hole when cooling is happening rapidly. In reality this cold gas should keep condensing to unresolved scales and form stars. This unphysical accumulation of cold gas is treated in previous work using different methods, including removing the cold gas with a sink term in the continuity equation \citep{Gaspari}, or using an effective equation of state appropriate for multiphase gas \citep{Dubois}. Since our simulation does not include these treatments, we will avoid deriving quantities that are sensitive to the central densities. For example, we will only use core-excised instead of total X-ray luminosity in later sections when we discuss cluster observables. 

Another thing to note is that any jet with nonzero thermal efficiency (run P5B and P5C) would produce a hot spot surrounding the black hole and hence a peak in the temperature profile near the center, which is also found by \citet{Gaspari11} and \citet{Dubois11}. Therefore, we advise simulators using concentrated thermal feedback such as thermal jets or quasar feedback to be cautious about this numerical effect when interpreting results near the region of feedback.    

%===================================================

%Emphasize core-excised quantities. Examine figure notations. 

\section{Implications for Cluster Observables}
\label{sec:observables}

\subsection{Robustness of Integrated Properties}

As seen in the previous section, specific models or parameters chosen can result in quite discrepant predictions for the evolution of the cluster profiles. Although the influence of AGN feedback is strongest in the core region, we may ask whether, under the influence of this feedback, the global cluster properties can still preserve the observed scalings. The robustness of integrated quantities (e.g.\ measured within $R_{500}$) is particularly crucial for cluster cosmology. Since current constraints are often derived using calibrations of the mass-observable relations informed by numerical simulations, it is necessary to quantify the systematic uncertainties due to incomplete knowledge of the details of AGN feedback processes.       

The first question we wish to address is whether any of the model variations explored in the previous section predict global cluster properties that violate the observed scaling relations. To this end we compute several observable quantities integrated within a sphere with radius $R_{500}$, including the gas mass $M_\mathrm{g}$, spectral-like temperature $T_\mathrm{sl}$ \citep{Mazzotta04}, X-ray luminosity $L_X$, integrated Compton $y$ parameter due to the Sunyaev-Zel'dovich (SZ; \citet{Sunyaev}) effect $Y_{SZ}$, and its X-ray analog $Y_X \equiv M_\mathrm{g}T_X$ \citep{Kravtsov}. Since the spectral-like temperature and the X-ray luminosity are very sensitive to dense and cold gas, we excised the core region ($< 0.15R_{500}$) in order to avoid numerical effects due to the cold gas accumulated around the SMBH as described in \S~\ref{sec:jet}. For run P4C ($R_0=5\ h^{-1}$\ kpc) and run P5A ($\epsilon_\mathrm{f}=0.02$, $\epsilon_\mathrm{m}=0\%$), the evolution after 9\ Gyr is not included because these runs encounter the cooling catastrophe. 

\begin{figure*}%[tp]
\begin{center}
\includegraphics[scale=0.5]{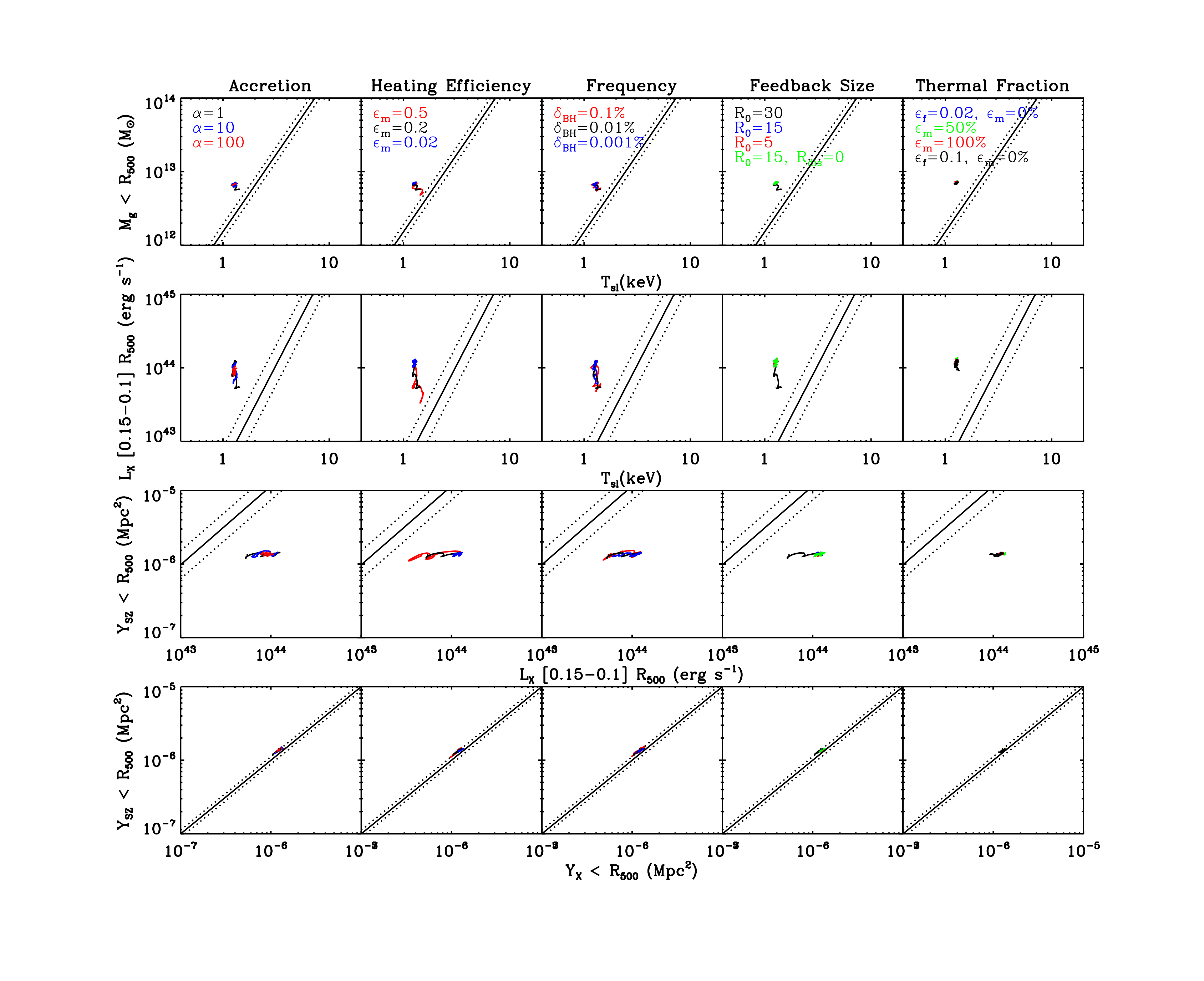} 
\caption{Trajectories of integrated observable properties on the scaling relations for variations of physical parameters explored in \S~\ref{sec:physical}. The rows from top to bottom show the $M_\mathrm{g}$--$T_\mathrm{sl}$, $L_X$--$T_\mathrm{sl}$, $Y_{SZ}$--$L_X$, and $Y_{SZ}$--$Y_X$ relations, respectively (see text for detailed definitions). The columns from left to right plot the runs with varied accretion model (group P1, Figure \ref{fig:accretion}), mechanical heating efficiency (group P2, Figure \ref{fig:effm}), frequency of feedback (group P3, Figure \ref{fig:dbh}), region of feedback (group P4, Figure \ref{fig:rbub}), and thermal to kinetic ratio (group P5, Figure \ref{fig:jets}), respectively. Overplotted are observed relations (solid) and r.m.s.\ scatter (dashed) for the REXCESS sample \citep{Croston, Pratt, Arnaud}. This figure illustrates that despite the discrepancies in the detailed evolution among different subgrid models as seen in previous figures, the integrated cluster observables still evolve with amplitudes that are consistent with the observed scatter. See Figure \ref{fig:obs_evol} and \ref{fig:mt_yy} for the detailed evolution of observables.}
\label{fig:scaling}
\end{center}
\end{figure*}

\begin{figure*}%[tp]
\begin{center}
\includegraphics[scale=0.5]{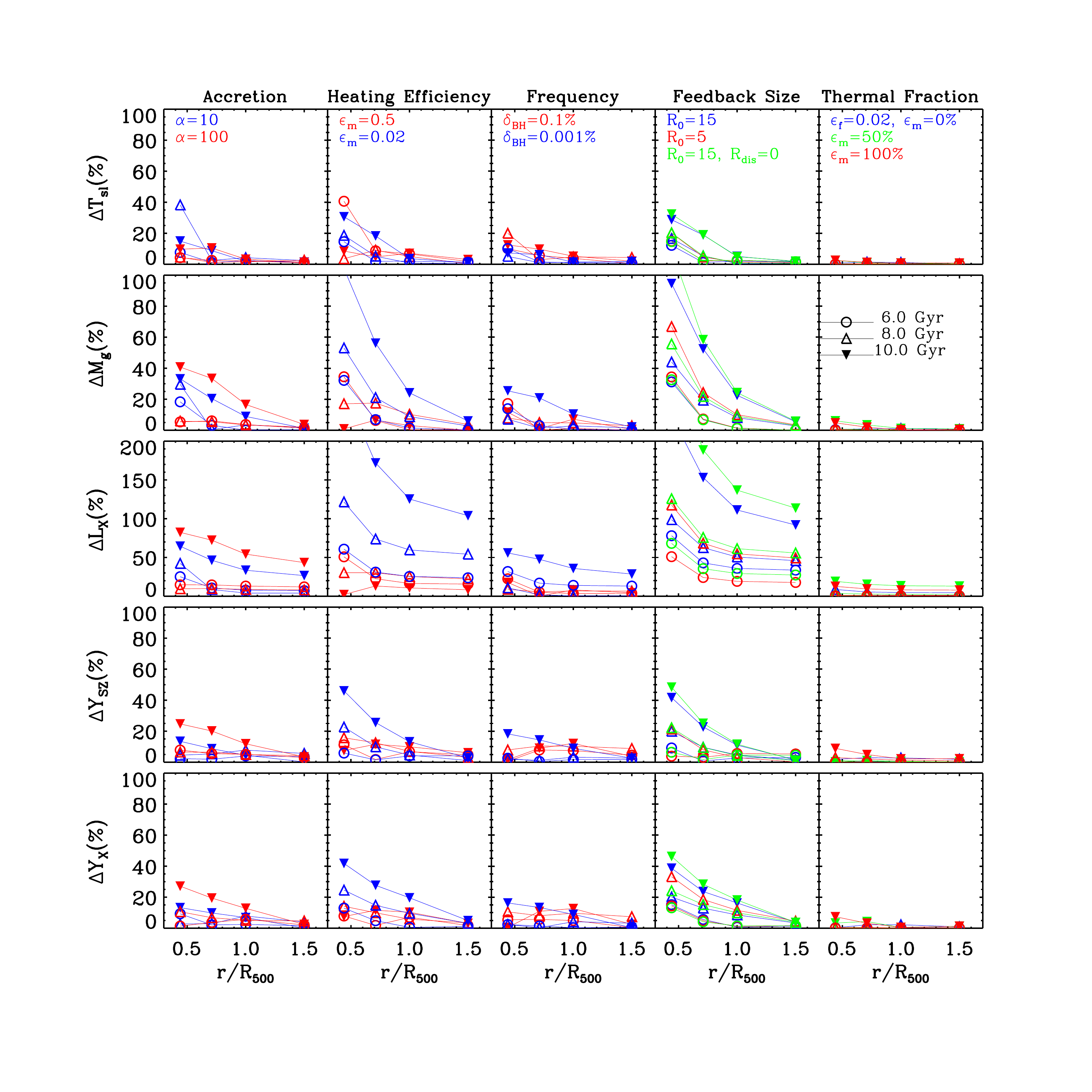} 
\caption{Uncertainties of cluster integrated properties due to AGN subgrid model variations (see Figure \ref{fig:scaling} for explanations of each column). Plotted are the predictions of $T_\mathrm{sl}$, $M_\mathrm{g}$, $L_X$, $Y_{SZ}$, and $Y_X$ (top to bottom; the core is excised for computing $T_{sl}$ and $L_X$; see text for details) relative to the fiducial runs as a function of four overdensity radii within which the observables are integrated, including $R_{2500}$, $R_{1000}$, $R_{500}$ and $R_{200}$. Each color corresponds to a specific run, and each line represents the result at a given simulation time. Note that the plotting range for the X-ray luminosity is $0-200\%$, since it has the largest uncertainty due to model variations. $T_\mathrm{sl}$, $Y_{SZ}$, and $Y_X$ are more robust. Comparing different groups of model variations (i.e.\ by columns), we find that the mechanical heating efficiency and size of feedback are the most influential, while feedback frequency and thermal to kinetic ratio play a minor role.} 
\label{fig:uncertainty}
\end{center}
\end{figure*}

Figure \ref{fig:scaling} compares the trajectories of observables to the scaling relations for the model variations explored in \S~\ref{sec:physical}. Each column compares results for a particular group in Table \ref{tbl:phys_par}, including variations in the accretion model, mechanical heating efficiency, feedback frequency, region of feedback, and thermal to kinetic ratio. From top to bottom we show the $M_\mathrm{g}$--$T_\mathrm{sl}$, $L_X$--$T_\mathrm{sl}$, $Y_{SZ}$--$L_X$, and $Y_{SZ}$--$Y_X$ relations and overplot the observed relations and scatter for the REXCESS sample \citep{Croston, Pratt, Arnaud}. We note that the offsets in the normalizations are due to the different ways to derive the {\it intrinsic} and {\it observed} quantities. Firstly, the observed values of $R_{500}$ in these studies are obtained by matching the empirical $M_{500}$--$Y_X$ relation in \citet{ArnaudMY}. As these authors point out, their hydrodynamic mass $M_{500}$ may underestimate the true mass. Therefore it is likely that their $R_{500}$ is smaller than that used in our computation, which lowers the values of $M_\mathrm{g}$ and $L_X$, but increases $T_\mathrm{sl}$, in a direction that could explain the shift in normalizations. Moreover, the X-ray luminosity in our calculation is bolometric and hence would be higher than the observed values, which are integrated over the energy range $0.1-2.4$\ keV. More detailed simulated observations are required for direct comparisons between the simulations and observations.

We find that despite the variation in the predicted cluster profiles produced by different subgrid models, the integrated properties for all the models evolve with amplitudes that are consistent with the observed scatter. In other words, when cooling is regulated, all the subgrid models are able to preserve global cluster properties as observed. This result gives us some confidence in the AGN subgrid models employed in cosmological simulations. 
However, it also implies that these various models and parameters cannot be distinguished by constraints on the integrated quantities of a single cluster alone, but must be constrained by observations that are more sensitive to cluster cores, or by comparing scaling relations of a sample of simulated clusters with observations.
%However, it also implies that these various models and parameters cannot be distinguished by constraints on the integrated quantities, but must be constrained by observations that are more sensitive to cluster cores.     

Although none of the individual trajectories in Figure \ref{fig:scaling} violates the observed scaling relations, when compared against each other, the predictions for a particular observable at a particular time can still vary significantly among models. Thus the second question to ask is: how large are the theoretical uncertainties due to different AGN subgrid models and variations in their parameters? Since AGN feedback is expected to have more impact at smaller radii, we compute the observables measured within several commonly-quoted overdensities, including $R_{2500}$, $R_{1000}$, $R_{500}$, and $R_{200}$, and compare their values with the fiducial runs, i.e., run P1A for the bubble model and run P4D for the jet model. The relative dispersion for a given observable $O$ is computed by $\Delta O \equiv |O-O_\mathrm{fiducial}|/O_\mathrm{fiducial}$. The results for five observables ($T_\mathrm{sl}$, $M_\mathrm{g}$, $L_X$, $Y_{SZ}$, and $Y_X$) with varied subgrid models are presented in Figure \ref{fig:uncertainty}.

Comparing the five observables, the X-ray luminosity has the largest uncertainties due to variations in subgrid models (note that its plotting range is $0-200\%$), and the gas mass is second. The other three variables ($T_\mathrm{sl}$, $Y_{SZ}$, and $Y_X$) are more robust. When different groups of model variations are contrasted, we find that the mechanical heating efficiency and the size of the feedback region cause the largest variations in the predicted observables. The influence of the accretion models and feedback frequency is smaller, and the thermal to kinetic ratio has the least impact. 

\begin{figure}%[tp]
\begin{center}
\includegraphics[scale=0.35]{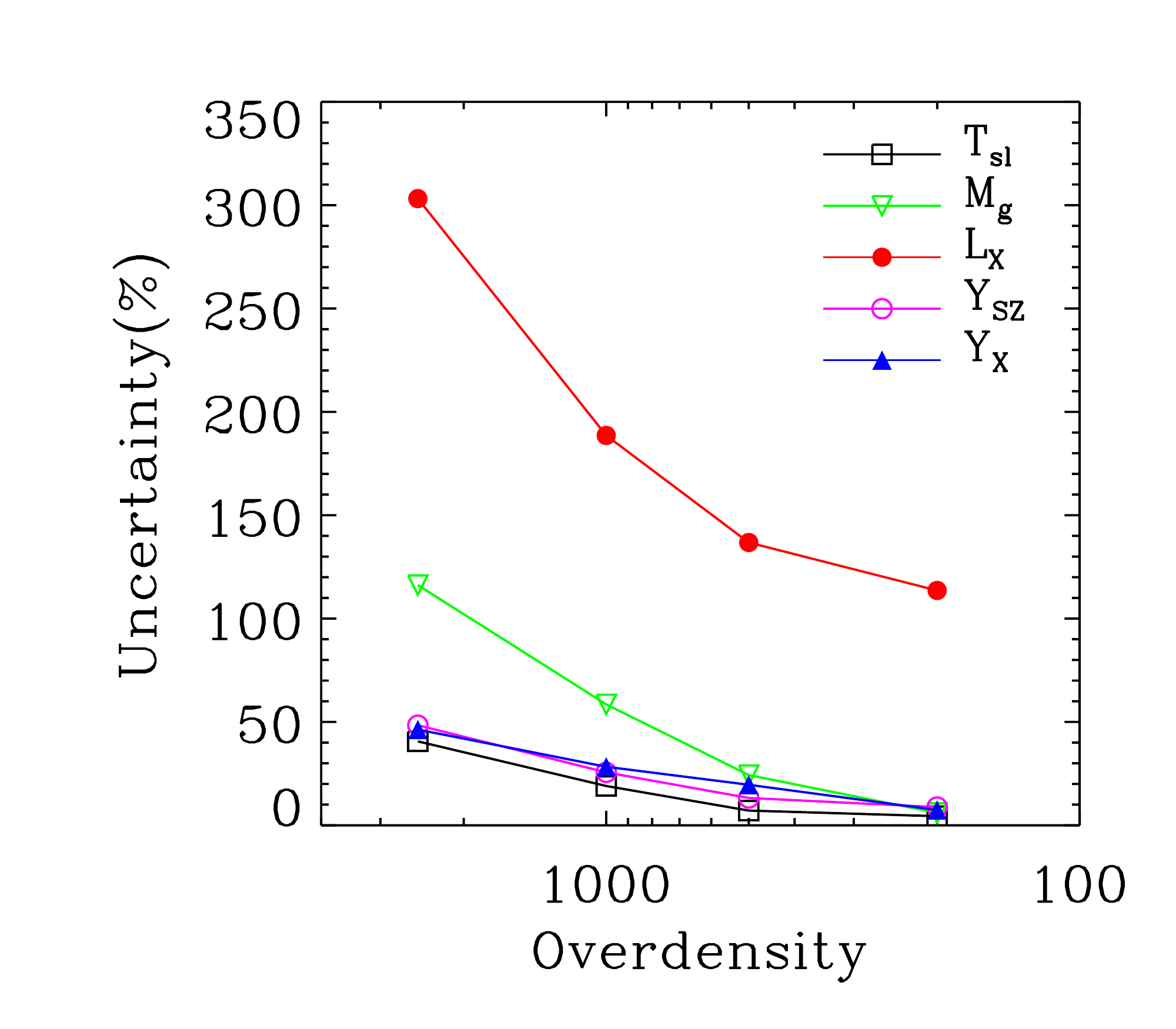} 
\caption{Model uncertainties for cluster observables as a function of overdensity radii. Values correspond to the maximum values across all models shown in Figure \ref{fig:uncertainty}.} 
\label{fig:uncertainty_summary}
\end{center}
\end{figure}

As expected, since feedback from the AGN is more influential toward the central SMBH, the model uncertainties are biggest for observables measured within $R_{2500}$. When quantities are integrated out to $R_{200}$, the uncertainties become small for all observables except the X-ray luminosity, because a large fraction of the total luminosity still comes from regions near the core. In Figure \ref{fig:uncertainty_summary} and Table \ref{tbl:uncertainty} we summarize the maximum uncertainties among all models for each cluster observable versus the overdensity radius. We find that in general observables that are sensitive to gas densities are more poorly predicted. The total gas mass predicted by different models has uncertainties ranging from a few percent at $R_{200}$, to $\sim20\%$ at $R_{500}$, to $\sim100\%$ at $R_{2500}$. Thus the X-ray luminosity, which is proportional to density squared, can vary by factors of a few for all radii. The level of uncertainties is smaller and comparable for $T_\mathrm{sl}$, $Y_{SZ}$, and $Y_X$, ranging from $\sim 40-50\%$ at $R_{2500}$, to $\sim 10-20\%$ at $R_{500}$, to $\sim 5-10\%$ at $R_{200}$. 

\subsection{Impact of AGN feedback on the Scaling Relations}

Keeping in mind the model uncertainties of integrated properties shown in the previous section due to different evolution in each model, next we study some general trends predicted by all the models. In particular, we probe the impact of AGN outbursts on the cluster observables using cross-correlations among them. Note that in this study we focus on the {\it global} observable properties, that is, whether there will be observable features beyond the core due to the disturbances introduced by the central AGN. The influence on the core properties by AGN has been discussed extensively in previous work (see \citet{McNamara} and references therein) and will be a part of our future work.  

\begin{table}
\caption{Model Uncertainties for Cluster Observables Measured within Various Overdensities.}  
\begin{center}
\begin{tabular}{cr@{.}lr@{.}lr@{.}lr@{.}l}
\hline
\hline
& \multicolumn{2}{c}{$\Delta = 2500$} &  \multicolumn{2}{c}{$1000$} &  \multicolumn{2}{c}{$500$} &  \multicolumn{2}{c}{$200$} \\ 
\hline
$\Delta T_\mathrm{sl}$ & 40 & 6\% & 19 & 0\% & 7 & 1\% & 4 & 4\% \\
$\Delta M_\mathrm{g}$ & 116 & 2\% & 58 & 5\% & 24 & 3\% & 6 & 0\% \\
$\Delta L_X$ & 303 & 2\% & 188 & 6\% & 136 & 8\% & 113 & 6\% \\
$\Delta Y_{SZ}$ & 48 & 4\% & 25 & 6\% & 13 & 2\% & 8 & 8\% \\
$\Delta Y_X$ & 46 & 3\% & 28 & 3\% & 19 & 6\% & 7 & 4\% \\
\hline
\hline
%\multicolumn{9}{l}{\footnotesize \footnotemark[1] Model uncertainties correspond to the maximum}\\
%\multicolumn{9}{l}{\footnotesize \footnotemark[0] values across all models shown in Figure \ref{fig:uncertainty}.}\\
\end{tabular}
\end{center}
\label{tbl:uncertainty}
\end{table}

\begin{figure*}
\begin{center}
\includegraphics[scale=0.65]{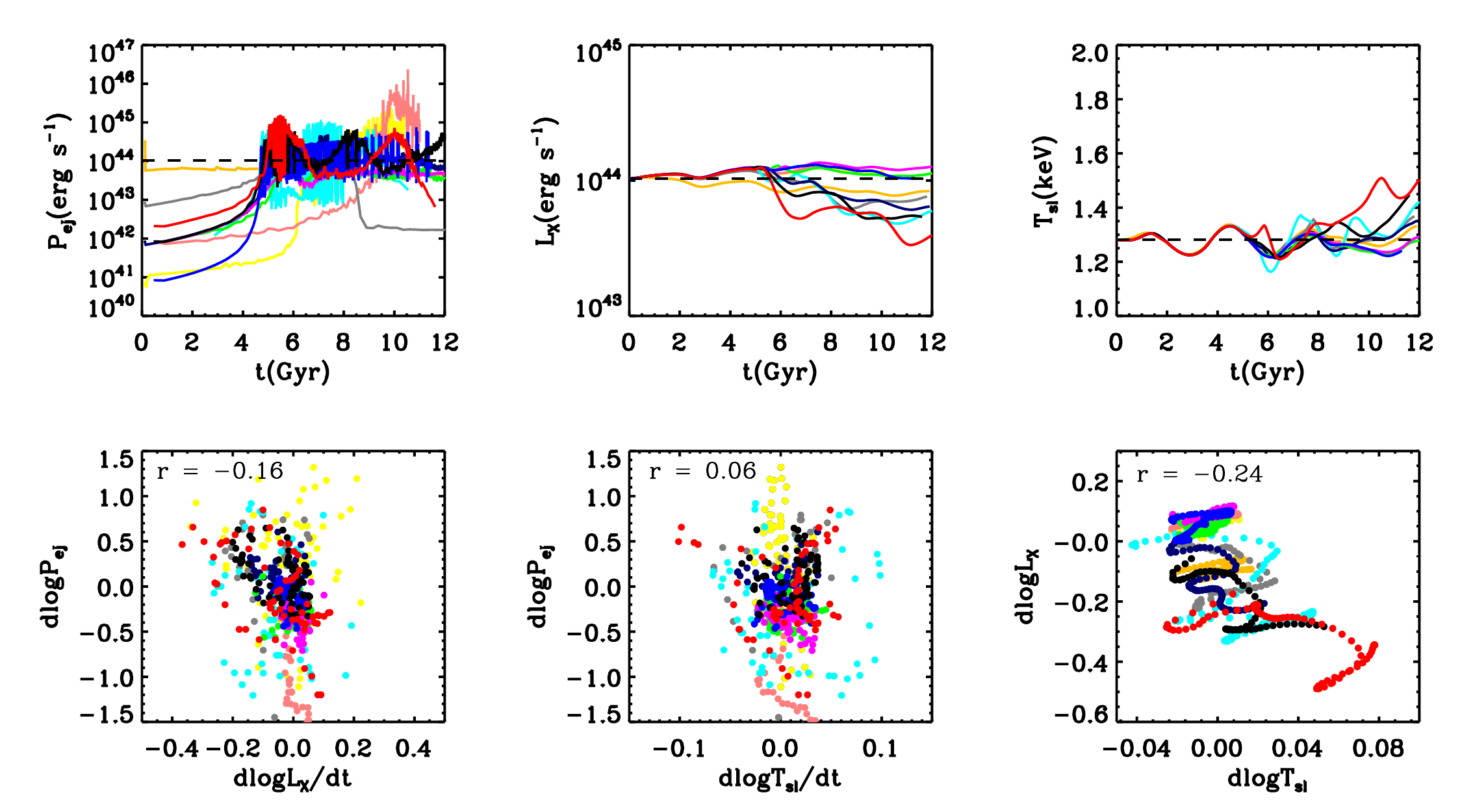} 
\caption{ {\it Top}: Evolution of the AGN power, X-ray luminosity, and spectral-like temperature measured within [0.15-1]$R_{500}$. All the bubble runs in Table \ref{tbl:phys_par} are shown in different colors ({\it black}: P1A, {\it gray}: P1B, {\it gold}: P1C, {\it blue}: P2A, {\it red}: P2B, {\it navy}: P3A, {\it cyan}: P3B, {\it green}: P4A, {\it pink}: P4B, {\it magenta}: P4C, {\it yellow}: P5E).   
{\it Bottom}: Correlations for data points in the time interval $6\leq t \leq12$\ Gyr ($r$ is the correlation coefficient) between the AGN power and the change in X-ray luminosity (left), the AGN power and the change in spectral-like temperature (middle), and the X-ray luminosity and spectral-like temperature (right). These correlations are expected because AGN outbursts result in reduction of the luminosity and heating of the cluster with slight time delays. } 
\label{fig:obs_evol}
\end{center}
\end{figure*}

Figure \ref{fig:obs_evol} (top row) plots the evolution of the AGN power, X-ray luminosity, and spectral-like temperature for all the bubble runs listed in Table \ref{tbl:phys_par}. The jet runs are not shown here because their AGN power cannot be compared directly in the thermal form. However, their results can be well represented by the jet-like bubble run P5E, as discussed in section \S~\ref{sec:jet} (see also Figure \ref{fig:bub+jet}). For all the models, the black hole self-regulates its growth when its feedback power is sufficient to balance the radiative losses by the cluster. However, the feedback power fluctuates around the mean after $t \simeq 6$\ Gyr with different amplitudes depending on the initial configurations and growth at earlier times. These fluctuations are present in the AGN activity as well as in the cluster observables. Taking run P2B (red curve) as an illustration, the strong AGN outbursts at $t \simeq 5-6$\ Gyr raise the entropy of the gas and hence induce the decrease in luminosity and increase in temperature at $t \simeq 6-7$\ Gyr. Similar effects can be seen for the peak in AGN activity around $t\simeq 10$\ Gyr and corresponding fluctuations in the observables at $t \simeq 11$\ Gyr.    

We therefore compare the AGN power with the {\it changes} in luminosity and temperature after $t=6$\ Gyr (bottom row in Figure \ref{fig:obs_evol}) to see whether their fluctuations are correlated. The notations dlog($L_{\rm X}$) and dlog($T_{\rm sl}$) represent logarithmic deviations from their initial values; the feedback power is plotted with respect to the initial luminosity too. The correlation coefficients are given by the Spearman Rank-Order Correlation test \citep{spearman}. As expected, there is a negative (positive) correlation between the feedback power and the time derivative of the luminosity (temperature). We note that the correlations are much weaker if the AGN power is compared with the {\it instantaneous} luminosity and temperature, because of the phase shifts between the peaks of AGN outbursts and the delayed responses of the ICM. Nevertheless, since the luminosity and temperature react to the feedback {\it in phase}, they have a strong anti-correlation as the system moves on the $L_X$--$T_\mathrm{sl}$ plane, as shown in the lower right panel in Figure \ref{fig:obs_evol}.

Recall that the ranges of trajectories on the $L_X$--$T_\mathrm{sl}$ plane predicted by all the AGN subgrid models are comparable to the observed scatter (Figure \ref{fig:scaling}). This implies that AGN feedback can drive a significant amount of the observed scatter in the $L_X$--$T_\mathrm{sl}$ relation, because of the anti-correlation between luminosity and temperature during the feedback events. This is in contrast to other physical processes such as cluster mergers, which tend to move the clusters along the scaling relations \citep{Yang}.

\begin{figure}
\begin{center}
\includegraphics[scale=0.58]{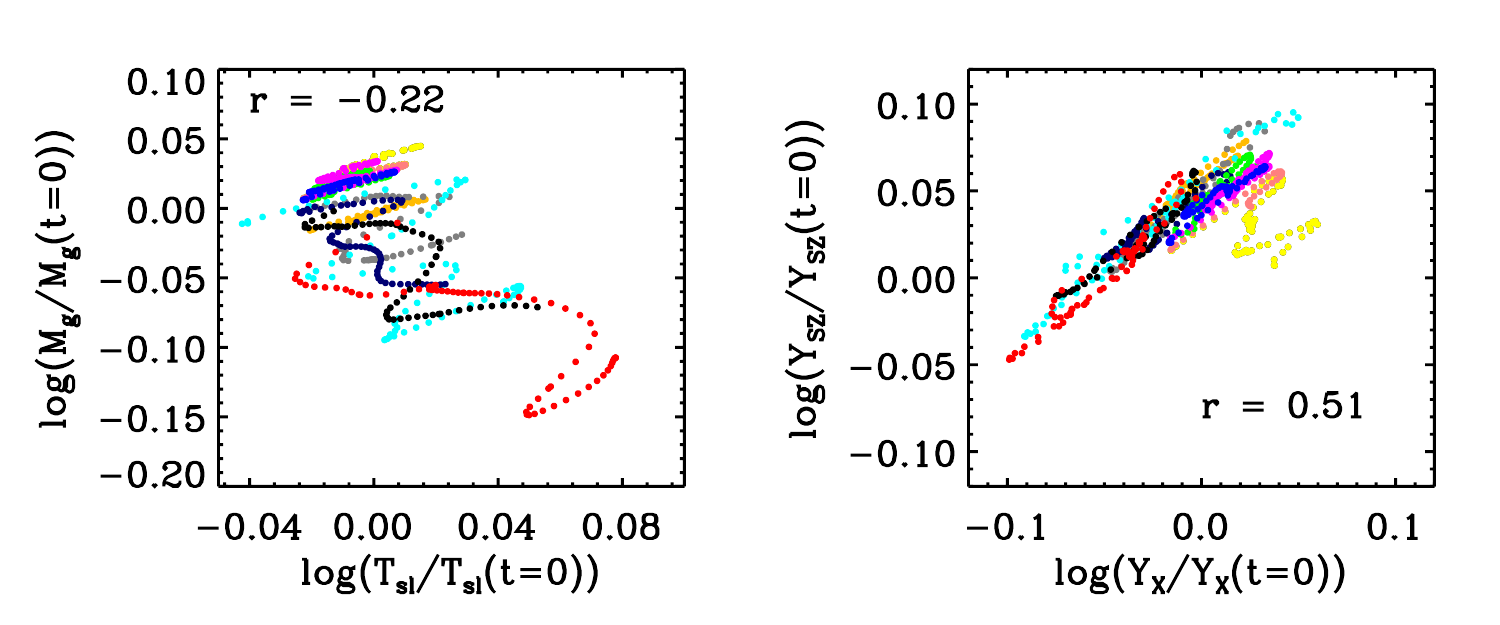} 
\caption{Trajectories on the $M_\mathrm{g}$--$T_\mathrm{sl}$ (left) and the $Y_{SZ}$--$Y_X$ (right) relations for $6\leq t \leq12$\ Gyr for the same runs as in Figure \ref{fig:obs_evol} ($r$ is the correlation coefficient). Similar to the $L_X$--$T_\mathrm{sl}$ relation, $M_\mathrm{g}$ and $T_\mathrm{sl}$ are anti-correlated. As a result, there is a tight positive correlation between $Y_{SZ}$ and $Y_X$.} 
\label{fig:mt_yy}
\end{center}
\end{figure}

Similarly, an anti-correlation exists between $M_\mathrm{g}$ and $T_\mathrm{sl}$ (Figure \ref{fig:mt_yy}, left panel), which may also contribute to the scatter in the $M_\mathrm{g}$--$T_\mathrm{sl}$ relation. 
Moreover, the anti-correlation implies that $Y_{SZ}$ and $Y_X$, which are essentially the products of $M_\mathrm{g}$ and $T_\mathrm{sl}$, do not deviate from the mean scaling relations significantly during AGN outbursts. As shown in the right panel of Figure \ref{fig:mt_yy}, the variation in both $Y$ parameters from the mid point is $\mbox{dlog}(Y)\sim 7.5\%$. Given the observed slope of the $Y$--$M$ relation of 1.82 \citep{ArnaudMY}, it corresponds to an implied uncertainty in mass $\mbox{dlog}M\sim 4.1\%$ if the $Y$--$M$ relation is used to get a cluster mass. This is smaller than the mass uncertainties inferred from other observables estimated by the same means, indicating that the $Y$ parameters are robust mass proxies even under the strong influence of energetic AGN outbursts.
%But at the same time, it also preserves the tight $Y_{SZ}$--$Y_X$ relation (right panel) since both $Y$ parameters are essentially the products of $M_\mathrm{g}$ and $T_\mathrm{sl}$. Therefore, we find that the $Y$ parameters are robust even under the strong influence of energetic AGN outbursts, which adds another reason why they are excellent tracers of cluster masses. 

\begin{figure*}%[tp]
\begin{center}
\includegraphics[scale=0.5]{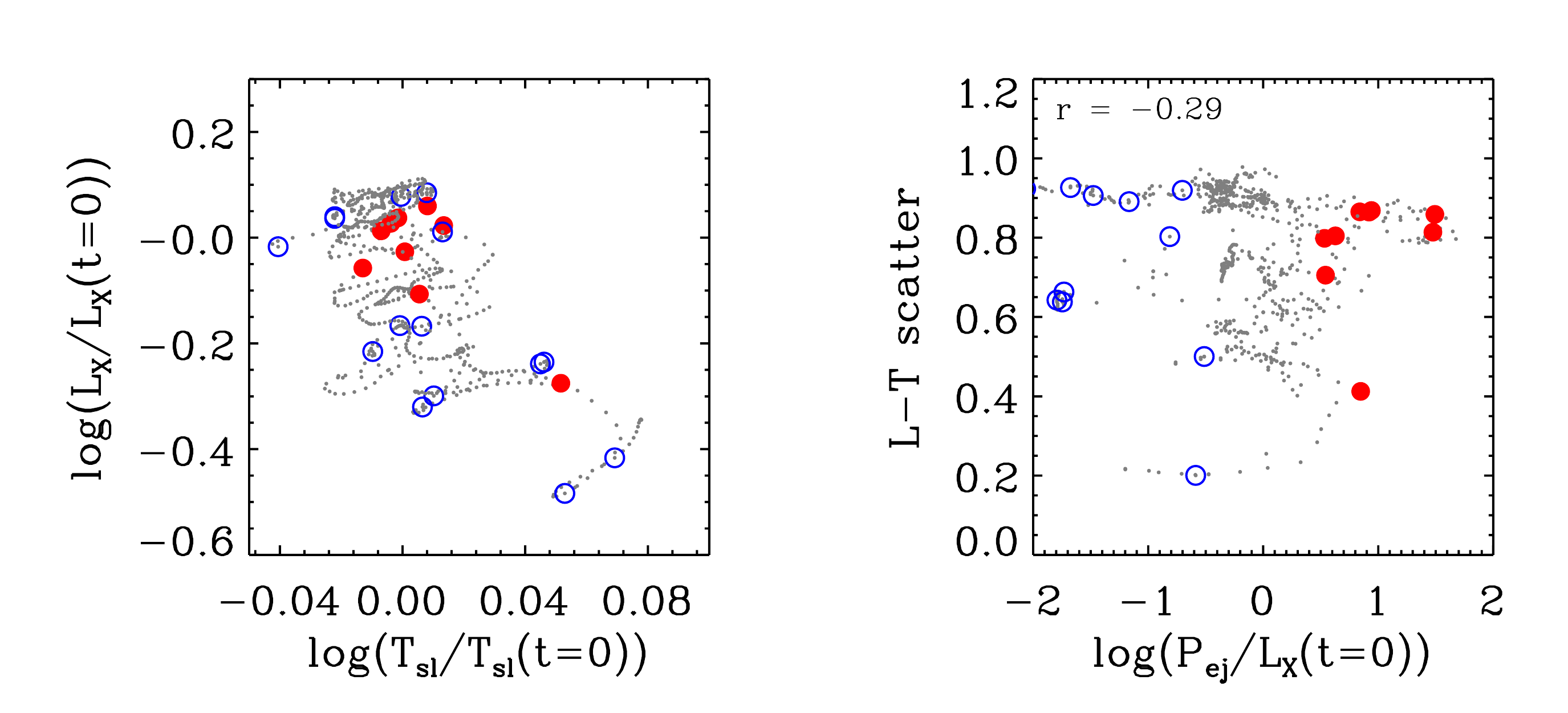} 
\caption{Trajectories on the $L_X$--$T_\mathrm{sl}$ relation (left) and correlation between its log scatter  and the power of AGN in the time interval $6\leq t \leq 12$\ Gyr for the same runs as in Figure \ref{fig:obs_evol} (right; $r$ is the correlation coefficient). Outbursts that are 0.5 dex more (less) powerful than the mean are marked in red filled circles (blue open circles). More powerful AGN preferentially have smaller scatter (lie below the mean); however, the correlation is diluted by the phase shift between the AGN power and the observables shown in Figure \ref{fig:obs_evol}.} 
\label{fig:lt_pej}
\end{center}
\end{figure*}

Since we have demonstrated that the scatter in the $L_X$--$T_\mathrm{sl}$ relation can be induced by feedback events, in Figure \ref{fig:lt_pej} (right panel) we correlate the $L_X$--$T_\mathrm{sl}$ scatter with the AGN power. Since we do not have a sample of clusters to derive the mean scaling relation, the scatter is computed by taking the logarithmic deviation from the observed relation shown in Figure \ref{fig:scaling}. As expected from the correlations found earlier (Figure \ref{fig:obs_evol}), a negative correlation exists between the scatter and the AGN power. However, again the trend is not prominent because of the phase shifts. 

This result may have implications for observational studies that attempt to connect the $L_X$--$T_\mathrm{sl}$ scatter to the AGN radio power. \citet{Croston05} found that radio loud AGN preferentially lie below the $L_X$--$T_X$ relation as evidence for AGN heating. However, a more recent study by \citet{Jetha} found a weaker relation. For illustration we plot the epochs when the AGN power is 0.5 dex more (less) than the zero-point value in filled (open) symbols (for clarity only data points at multiples of one Gyr are shown). As can be seen in the left panel of Figure \ref{fig:lt_pej}, there is no clear segregation on the $L_X$--$T_\mathrm{sl}$ plane between the more powerful and the more quiescent populations, because the correlation between the scatter and AGN power is not strong enough (right panel). If the radio loudness is (roughly) proportional to the power of AGN, this may explain why observationally it is difficult to find a strong correspondence for a sample of clusters.

\begin{figure*}%[tp]
\begin{center}
\includegraphics[scale=0.5]{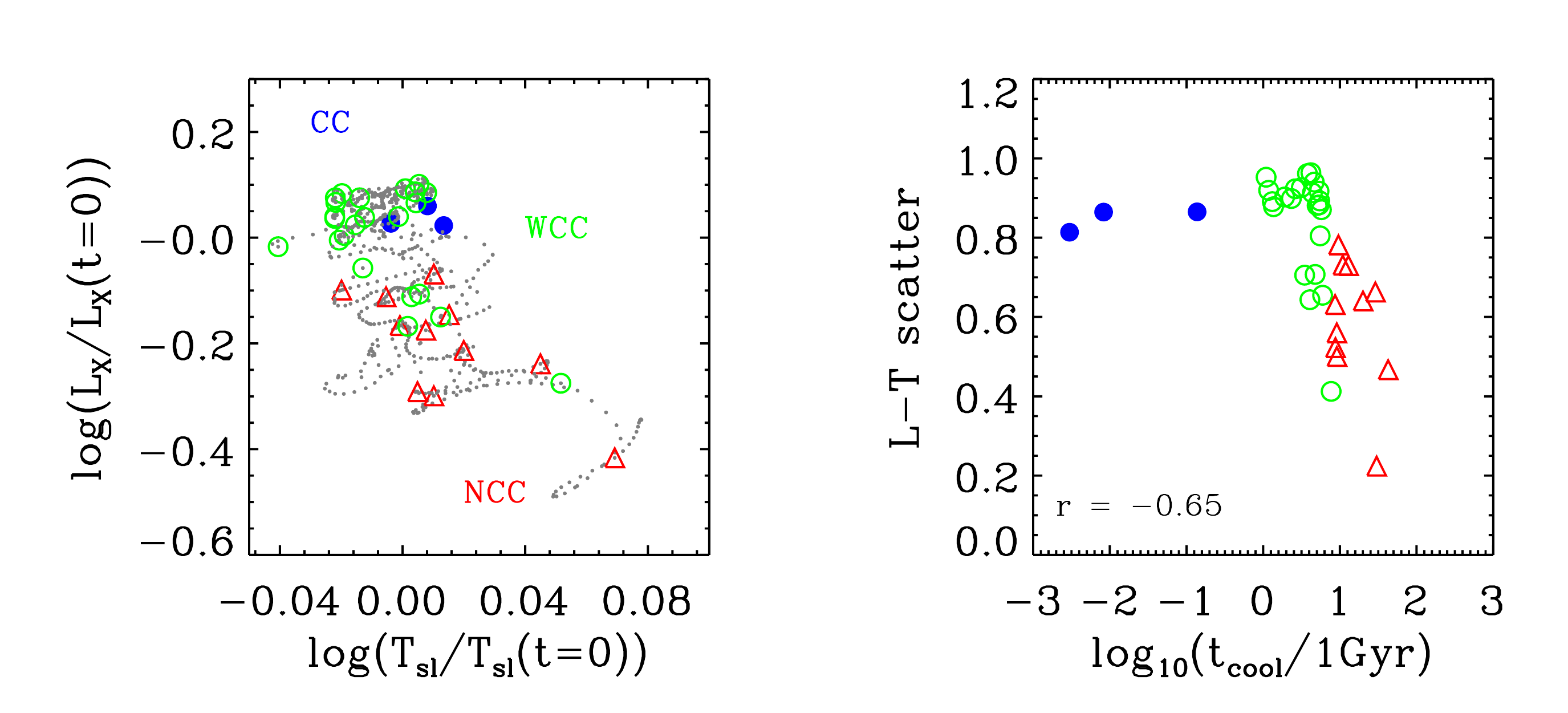} 
\caption{Trajectories on the $L_X$--$T_\mathrm{sl}$ relation (left) and correlation of its long scatter with the cooling time in the cluster core in the interval $6\leq t\leq12$\ Gyr for runs shown in Figure \ref{fig:obs_evol} (right; $r$ is the correlation coefficient). Clusters with CC, WCC, and NCC are plotted with blue filled circles, green open circles, and red open triangles, respectively. There is a clear anti-correlation between the $L_X$--$T_\mathrm{sl}$ scatter and the cooling time, so that CC clusters tend to lie above the mean relation.} 
\label{fig:lt_tcool}
\end{center}
\end{figure*}

The scatter in the $L_X$--$T_X$ relation has long been known to be dominated by the core properties of clusters; excluding the emission inside the core region can significantly reduce the scatter \citep[e.g.][]{Pratt}. Moreover, CC clusters generally have a higher normalization on the plane than NCC clusters. Here we explicitly show in Figure \ref{fig:lt_tcool} that such a trend can be caused by the effects of AGN. During the AGN feedback events, the $L_X$--$T_\mathrm{sl}$ scatter is anti-correlated with the cooling time (right panel). Thus the CC and WCC clusters tend to lie on the upper half of the relation compared to NCC clusters (left panel). Since the luminosity and temperature studied here are core-excluded, even stronger trends are expected to be found for the core-included $L_X$--$T_{\rm sl}$ relation. 

Note however that the NCC clusters at later times in the simulations are mostly produced by models in which the AGN feedback is either very powerful (high mechanical heating efficiency) or very extended (large bubble sizes), whereas the CC clusters are produced by the jet models. Therefore the exact amplitude of this segregation of CC and NCC clusters, or the suppression of $L_X$--$T_{sl}$ normalization, would depend upon model selection. Interestingly, cosmological simulations including bubble feedback have shown that AGN feedback is capable of suppressing the $L_X$--$T_{sl}$ normalization for low-mass clusters and steepening the slope to match observations \citep{Puchwein08}. However, our study suggests that there can be systematically different results if one chooses different models, such as models with smaller mechanical heating efficiency, or the jet models.   

%===================================================

\section{Discussion and Conclusions}
\label{sec:conclusion}

Feedback from the AGN is a crucial ingredient in modeling the observable properties of galaxy clusters.
In the literature there has been a variety of AGN subgrid models employed in cosmological simulations. However, systematic parameter surveys and comparisons among different implementations are critical for understanding the robustness of their predictions. In this study, we implemented several commonly-adopted accretion and feedback models into FLASH and systematically explored various parameters in an idealized cluster atmosphere. We first performed a sensitivity test of these subgrid models using a spectrum of parameters in order to understand their relative importance. We then quantified the theoretical uncertainties of cluster integrated properties due to model variations, and studied the impact of AGN feedback on the scaling relations by summarizing the results among different models. 

Since it is infeasible to explore every implementation and parameter of existing AGN subgrid models, our study is focused on two common approaches: to inject thermal energy in an extended region to mimic already inflated bubbles \citep{Sijacki}, and to inject mass and momentum as well in the form of bipolar jets \citep{Cattaneo}. For the survey of parameter sensitivity, we investigated parameters that are least constrained by observations, including both numerically relevant parameters (Table \ref{tbl:num_bub_par} and Table \ref{tbl:num_jet_par}) and physically motivated parameters (Table \ref{tbl:phys_par}). We compared their influence on the evolution of SMBH and cluster properties and examined their ability to self-regulate and reproduce observed profiles inside cluster cores. The main findings of the sensitivity study are summarized in the following. 

1.\ {\it Resolution -} The convergence tests show that increasing resolution generally produces more variable accretion rates. The bubble feedback suffers greater variation by changing the resolution, whereas the jet model is more robust, as long as the radius for computing accretion rates is larger than the sizes of the jets.  

2.\ {\it Accretion -} The proportionality used to relate the Bondi accretion to the actual SMBH accretion rate has a significant impact on the evolution of SMBH and cluster properties. Given the uncertainties in the accretion mechanisms, current AGN subgrid models may have very limited power to predict the evolution of cluster core properties, such as the fraction of CC versus NCC clusters as a function of time. 

3.\ {\it Efficiency of mechanical heating -} Varying the mechanical heating efficiency does not alter the overall evolution as much as accretion. Feedback with large efficiencies has more variable accretion rates, more suppression in black hole growth, and higher entropy floors. Efficiencies that are too small would fail to overcome cooling. 

4. {\it Frequency -} Changing the frequencies of injections has a minor effect. Longer duty cycles tend to generate more fluctuations in the accretion rates and cluster profiles.

5. {\it Region -} The evolution of the SMBH and the ICM is very sensitive to the size of the energy injection region (the displacement from the BH does not matter much). Moreover, bubbles that are too large would sometimes produce entropy floors that are inconsistent with observations, and bubbles that are too small may not be able to heat the entire CC and stop catastrophic cooling. Thus for any model that requires setting the feedback sizes by hand, there would be an issue of fine-tuning for a general population of clusters. 

6. {\it Kinetic feedback -} The jets with varied thermal to kinetic ratios produce very similar results. A combination of thermal and kinetic energy is slightly more efficient than purely thermal feedback. Purely kinetic feedback with efficiencies that are too small would fail to self-regulate. 

Comparing the bubble and jet models, we find that their main difference lies in the sizes of energy injection regions (Figure \ref{fig:bub+jet}). The models are numerically degenerate when appropriate parameters are chosen, i.e., producing tiny, continuous bubbles to mimic the jets. The jet model is in general more robust to many numerical parameters (e.g.\ resolution) as well as physical parameters (e.g.\ sizes of feedback, which is desirable because there is no need for fine-tuning). However, though the jets can maintain the cluster in the CC state, to avoid the artificial accumulation of cold gas around the black hole requires better treatment of the multiphase gas. Also, purely thermal concentrated heating, like thermal jets or quasar feedback, would produce a central peak in the temperature profile. Therefore, one needs to be cautious when interpreting results in the immediate surroundings of the BH. 

%Systematic difference in AGN power by using bubble or jet models.

Outbursts from AGN are energetic events that can greatly influence the observable properties of galaxy clusters. Previous simulations with AGN subgrid models have either studied their impact inside cluster cores \citep{Gaspari, Dubois}, or focused on matching the observed scalings of SMBH evolution and cluster gross properties \citep{Sijacki, Booth, Puchwein08}.  However, whether these models can simultaneously reproduce cluster properties both inside and outside the cores has not previously been demonstrated. In the sensitivity study we identified model parameters that can self-regulate and produce core profiles consistent with observations. Now we summarize our findings for cluster integrated properties as follows.

1.\ All the subgrid models that successfully regulate cooling in the previous analysis also produce variation in integrated quantities consistent with the scatter of the observed scaling relations (Figure \ref{fig:scaling}).

2.\ The model uncertainties in $M_{\rm g}$, $T_{\rm sl}$, $L_X$, $Y_{SZ}$, and $Y_X$ as functions of overdensity radius are quantified in Table \ref{tbl:uncertainty}. Quantities that are more sensitive to gas density (e.g.\ $M_{\rm g}$, $L_X$) have larger uncertainties, whereas $T_{\rm sl}$, $Y_{SZ}$, and $Y_X$ are most robust to model variations, to the levels of $\sim 10-20\%$ at $R_{500}$, and $\sim 5-10\%$ at $R_{200}$. 

3.\ Since AGN feedback reduces gas density and raises temperature, anti-correlations exist between $L_X$ and $T_{\rm sl}$ and also between $M_{\rm g}$ and $T_{\rm sl}$, contributing to the intrinsic scatter in these two scaling relations. However, because the ICM reacts to AGN feedback with a delay, correlations between observables and AGN power are weak.

4.\ Because $M_{\rm g}$ and $T_{\rm sl}$ are anti-correlated, even under the influence of strong AGN outbursts, the $Y_{SZ}$ and $Y_X$ parameters are still robust mass proxies. 

Contrasting the bubble and jet models, we find that the more extended bubble injections are generally more effective in altering cluster properties than the more concentrated jet feedback. Consequently, simulations using the bubble feedback model and the accretion strength $\alpha=100$ \citep{Puchwein08} are able to steepen the $L_X$--$T_X$ slope but have difficulties producing CC clusters as observed. Though studies based on an improved accretion model (i.e.\ the $\beta$ model proposed by \cite{Booth}) have successfully matched the $L_X$--$T_X$ slope and other properties on group scales \citep{McCarthy10}, future simulations on the cluster scale are required to verify whether CC clusters can be produced. If not, then it could mean that the sizes of the bubbles are still too large and have to be further controlled, or that other accretion models need to be considered. On the other hand, though simulations adopting the jet model (either for an idealized cluster as in our study, or re-simulations from a cosmological volume as in \cite{Dubois} and \cite{Gaspari}) can successfully maintain clusters in the CC state, a statistical sample of clusters generated from a full cosmological simulation using the jet model does not yet exist to verify whether the jets could provide enough entropy to steepen the $L_X$--$T_X$ relation. If not, either the feedback energy needs to be distributed by some other mechanism, or the solution still lies in other accretion models. We recommend that these possibilities should be investigated to understand the limitations of the existing bubble and jet models before one attempts to refine the parameter space of any particular model.

The integrated Compton $y$ parameter, $Y_{SZ}$, and its X-ray analog, $Y_X$, are considered very good cluster mass proxies because previous simulations (without AGN feedback) show that they have very small mass scatter \citep{Kravtsov} and they are relatively insensitive to cluster dynamical state \citep{Poole, Wik08, Yang}. Here we further show that the $Y$ parameters are not easily disturbed by powerful AGN outbursts, which adds another reason to why they present so little observed scatter and can be used as excellent mass tracers.  

As it becomes more common to use the scaling relations of the $Y$ parameters provided by numerical simulations for calibration in observational studies \citep{Arnaud} or for deriving cosmological constraints \citep{Mantz08, Vikhlinin, Vanderlinde}, we note that incomplete knowledge of the processes of AGN feedback puts a limit on the predictive power of current cosmological simulations. Even for these most robust variables, the theoretical uncertainties due to model variations are $\sim 10-20\%$ at $R_{500}$ and $\sim 5-10\%$ at $R_{200}$, which would translate into mass errors comparable to other main sources of systematic errors reported in the literature, such as the bias of hydrostatic mass due to non-thermal pressure support \citep[e.g.][]{Lau}. Those variables that are sensitive to gas density (e.g.\ $L_X$ in particular) are even more uncertain. 
We note that the level of model uncertainties may be dependent on cluster masses, as the gas would be more easily displaced by AGN feedback in lower mass systems. For clusters more massive than our simulated cluster, the model uncertainties of integrated quantities may be smaller than the values quoted above. However, since the predictions get progressively worse near cluster cores, it is very likely that there are still substantial uncertainties inside the cores of more massive clusters. Furthermore, since clusters form hierarchically, the core properties (e.g.\ entropy floor) of massive clusters at the present day are determined by when and how the gas is heated inside the lower-mass systems that later merge into the clusters \citep[e.g.][]{McCarthy11}.  
Therefore, in order to improve predictions of cluster observables (including the cores) and derive cosmological constraints to the percent level, it is essential for numerical simulations to focus on how to improve the modeling of the subgrid physics before making various predictions.

Our sensitivity study showed that in order to effectively improve future AGN subgrid models, the crucial next step is to further constrain the accretion processes of the SMBH, the mechanical heating efficiency, and the sizes of the feedback region. The mechanical heating efficiency parameter (i.e.\ the fraction of feedback energy transformed into heat) and the sizes of the region to distribute heat may be evaluated from detailed numerical simulations \citep{Vernaleo, ONeill10}. However, as discussed in \S~\ref{sec:effm}, the results would depend on physics included in the simulations. Thus to pin down these parameters still requires more knowledge of the mixing properties of the ICM. For the mechanical heating efficiency, before its value can be estimated reliably, an alternative way is to adjust the efficiency to match the normalization of the $M_\mathrm{BH}$--$\sigma$ relation or the cosmic black hole mass density at $z=0$ \citep{Sijacki, Booth, Dubois11}. Note that since the obtained value is also dependent upon the specific accretion and feedback model employed, the efficiency parameter, if determined in this fashion, cannot inherit from other simulations but will have to be normalized for each realization. 

Improving the subgrid accretion model is a more challenging task, simply because it is still unclear how to link the accretion rates across such a great dynamical range. So far most cosmological simulations evaluate accretion rates based on the Bondi accretion rate. While the Bondi accretion rate appears sufficient in powering the observed AGN jets for many cases \citep{Allen}, for some systems other mechanisms seem to be required \citep{McNamara11}. Furthermore, the original Bondi accretion rate is based on simplified assumptions such as spherically-symmetric, steady flow with zero velocity at the Bondi radius. These criteria may not be applicable to all systems, especially those with radial infall due to rapid cooling, or with non-negligible angular momentum \citep[e.g.][]{Power11}. Alternative schemes have been proposed, including stochastic accretion \citep{Pope07}, cold gas accretion \citep{Pizzolato05}, accretion by gravitational instabilities in galaxies \citep{Hopkins10}, and SMBH spins \citep{McNamara09}. These models are not yet integrated into cosmological simulations (except a cold-accretion-like scheme used in \citet{Gaspari}). Clearly more detailed investigations and comparisons in this area are necessary for further improvement of the AGN subgrid models.

%===================================================

\section*{Acknowledgments}

HYY acknowledges support from a NASA Earth and Space Science Fellowship (NNX08AZ02H). PMS acknowledges support from a DOE Computational Science Graduate Fellowship (DEFG02-97ER25308). We acknowledge support under a Presidential Early Career Award from the U.S. Department of Energy, Lawrence Livermore National Laboratory (contract B532720) and from NASA (grant NNX06AG57G). Resources supporting this work were provided by the NASA High-End Computing (HEC) Program through the NASA Advanced Supercomputing (NAS) Division at Ames Research Center. FLASH was developed largely by the DOE-supported ASC/Alliances Center for Astrophysical Thermonuclear Flashes at the University of Chicago.

%===================================================

\bibliography{agn}

\label{lastpage}

\end{document}